\documentclass{elsart}
\usepackage[dvips]{graphics}
\usepackage{amsfonts}
\usepackage{amsmath}
\usepackage{ams}
\usepackage{amssymb}
\newcommand{\Pb}{\mathbb{P}}
\newcommand{\E}{\mathbb{E}}
\newcommand{\reals}{\mathbb{R}}
\newcommand{\pc}{\mathbb{P}}

\newcommand{\bx}{\mathbf{x}}
\newcommand{\by}{\mathbf{y}}
\newcommand{\bz}{\mathbf{z}}
\newcommand{\bp}{\mathbf{p}}
\newcommand{\bq}{\mathbf{q}}
\newcommand{\bk}{\mathbf{k}}
\newcommand{\ket}[1]{\left.|#1\right>}
\newcommand{\bra}[1]{\left<#1|\right.}
\newcommand{\bracket}[2]{\left<#1|#2\right>}

\begin{document}
\begin{frontmatter}
\title{Product structure of heat phase space\\ and Branching Brownian motion}

\author{Frederic P. Schuller}
\address{Department of Applied Mathematics and Theoretical Physics, 
\\Centre for Mathematical Sciences, University of Cambridge\\
Cambridge CB3 0WA, United Kingdom}
\ead{F.P.Schuller@damtp.cam.ac.uk}
\author{Pascal Vogt}
\address{Department of Mathematical Sciences,
University of Bath,\\ 
Bath BA2 7AY, United Kingdom}
\ead{P.Vogt@maths.bath.ac.uk}

\begin{abstract}
A generical formalism for the discussion of Brownian processes with 
non-constant particle number is developed, based on the
observation that the phase space of heat possesses a product structure
that can be encoded in a commutative unit ring. 
A single Brownian particle is discussed in a Hilbert module theory,
with the underlying ring structure seen to be intimately linked to the
non-differentiability of Brownian paths.
Multi-particle systems with interactions are explicitly
constructed using a Fock space approach. 
The resulting ring-valued quantum field theory is applied to binary
branching Brownian motion, whose Dyson-Schwinger equations
can be exactly solved. 
The presented formalism permits the application of the full machinery of
quantum field theory to Brownian processes.
\end{abstract}


\end{frontmatter}

\newpage
\section{Introduction}
The study of Brownian motion, and Markov processes based on it, has
been a rapidly growing subject in both mathematics and physics since
Einstein's pioneering paper \cite{Einstein}, linking the macroscopic heat flow
to microscopic Brownian motion. The rigorous probability theoretical
formulation is due to Wiener \cite{Wiener}, leading to a well-defined Feynman-Kac path
integral representation for solutions of the heat equation. 
Remarkable non-trivial results, as the non-differentiability
of Brownian paths \cite{Paley}, follow from Wiener's
construction. Today, Brownian motion is a very active area of research
in both pure mathematics and physics \cite{INW} \cite{Dynkin} \cite{Dawson}, as well as in applications to a
wide variety of phenomena in nature and finance \cite{Karatzas}.

Multi-particle Brownian systems with dynamical creation
and annihilation of particles, such as branching and dying processes, 
are conventionally studied by non-linear extensions
of the heat equation. Such a treatment, however, requires 
tailor-made kinematics for each individual system at hand, in order to      
encode an intrinsically multi-particulate problem in the formalism
of a single heat equation. This method becomes almost prohibitively difficult
in application to the currently intensively studied catalytic
processes \cite{catalytic}.  The aim of this paper is to remedy this
situation and to devise a
generical kinematical framework for \textsl{any} Brownian process featuring a
dynamically changing particle number.\\ 

Our investigation starts with the crucial observation that the phase space of 
the heat equation carries a product structure, in
contrast to the complex structure of the phase space for the
Schr\"odinger equation. The same way in which the latter gives rise to
the ubiquitous occurrence of the complex numbers in quantum theory,
the product structure can be absorbed into the commutative unit ring
$\pc$ of pseudo-complex numbers, governing the kinematics of Brownian
particles. 
The study of one-particle Brownian
motion then becomes a Hilbert module theory over $\pc$, where the
non-differentiability of Brownian paths emerges as a direct
consequence of the pseudo-complex structure of heat phase space. This
thorough geometrical understanding of the one-particle case can then be
put to use in the construction of multi-particle systems, employing a
Fock space approach. 
 In contrast to the conventional treatment, the presented pseudo-complex
Hilbert module description of the single particle
is extended to a \textsl{model-independent} Fock space formalism, and the dynamics of
an operator-valued second quantized field are derived. In this
set-up, we discuss general interacting pseudo-complex quantum field
theories as the appropriate kinematical framework for the discussion
of Brownian processes with non-constant particle number.

In general, the advantage of this field theoretical point of view is
that one can easily write down models for
arbitrarily complex Brownian processes. Ideally, these models have
exactly solvable Dyson-Schwinger equations.
Otherwise, a truncated series expansion of interesting quantities in
terms of Feynman diagrams may be just the appropriate tool to extract
information on analytically inaccessible systems. 

The abstract formalism is applied to the classically
well-studied binary branching Brownian motion. The Dyson-Schwinger
equations of this system can be exactly solved, and coincide with the
equations of motion that are found classically for corresponding
quantities. A point in case is the reproduction of the extinction
probability for such a process, which appears as the dressed one-point
function in the pseudo-complex quantum field theory. The formalism is
versatile enough to embrace models with dynamical catalysts, which
currently receive much attention \cite{catalytic}. 

The organization of the paper is as follows. Section II reviews the
conventional treatment of multi-particle Brownian processes for the
benefit of the non-specialist reader. Section III identifies the product
structure of the heat phase space, and introduces the pseudo-complex
ring. Section IV discusses a single Brownian particle in a
pseudo-complex Hilbert module formalism, preparing the non-interacting
multi-particle theory developed in Section V. Section VI studies theories
with dynamical particle creation and annihilation. These abstract
developments are applied to the binary branching model and a
self-intoxicating particle in Section VII. We
summarize and conclude in Section VIII.

\section{A review of Brownian processes}\label{sec_review}
The Brownian motion of a single particle is well-understood and
can be given a rigorous probability theoretical description. More
complex systems, such as spatial branching processes, are of current research
interest \cite{INW} \cite{Dynkin} \cite{Dawson} \cite{Etheridge}. They are conventionally studied as extensions of the
single-particle problem, but the kinematical framework has to be
adapted  from case to case. This section briefly reviews some of the
well-studied processes derived from Brownian motion for
the non-specialist reader, and closes with a critical discussion. In
particular, we will argue that a general framework for the study of
systems with non-constant particle number is desirable, such that a system is
completely specified by its dynamical equations, without the need to adapt 
the underlying concepts. The construction of such a system-independent
kinematical framework for Brownian processes is the aim of this paper.

\subsection{Heat equation and free Brownian motion}
Recall that the {\em heat equation} is given by
\begin{equation}
  \partial_t \, \varphi(t,\bx) = \frac{1}{2} \Delta \, \varphi(t,\bx),
  \label{heat}
\end{equation}
where $\varphi$ is a real-valued field on non-relativistic space time
$\reals^0_+ \times \reals^d$, and the diffusion constant is set to
unity.
Given the initial condition $\varphi(0,\bx)=u(\bx)$, the linear partial 
differential equation (\ref{heat})  has a unique solution. Its physical significance is the following: suppose the function $\varphi(t,\bx)$ 
represents the temperature at time $t$ and position $x$, then starting with the 
initial temperature distribution $u$, the heat equation 
describes the heat flow in time. In this sense, the heat equation governs the
{\em macroscopic} behaviour of the temperature field.

In 1905, Einstein
discovered the relation between the heat equation and a stochastic
process \cite{Einstein}
called {\em Brownian motion}. Intuitively, this process describes the
{\em microscopic} behaviour of a \textsl{single particle} that contributes to
the heat.
Mathematically speaking, a Brownian motion with start in $x \in \reals^d$
is a stochastic process, i.e., a random path
$\{B_t : t \geq 0\}$ which is continuous and has independent, stationary and normally
distributed increments. A Brownian motion can therefore be considered,
in a rigorous probabilistic manner \cite{Wiener}, as a random element of the 
canonical Wiener space $(C([0,\infty)), \mathcal{F}, \Pb_x)$, where 
$\Pb_x$ is a measure on the space
$(C([0,\infty),\mathcal{F})$ of continuous paths in $\reals^d$ equipped
with the $\sigma$-field $\mathcal{F}$ which is generated by the family of
projections $\pi_t : C[0,\infty) \rightarrow \mathbb{R}^d$,
$ \omega \mapsto \omega_t$. The Wiener measure
$\Pb_x$ satisfies, for all times $t_1, t_2, \dots,
t_n$ and all Borel sets $A_1,A_2,\dots , A_n \subseteq \reals^d$, the relation
\begin{eqnarray}
  \lefteqn{\Pb_x \{ B_{t_1} \in A_1, B_{t_2} \in A_2, \dots , B_{t_n} \in A_n 
  \}} \nonumber \\  
  &=& \int_{A_1} \int_{A_2} \dots \int_{A_n}
    d^dx_1\dots d^dx_n \, 
    p_{t_1} (\bx,\bx_1) \dots p_{t_n-t_{n-1}}(\bx_{n-1},\bx_n),
\end{eqnarray}
where the transition density of Brownian motion is given by the heat kernel
\begin{equation}\label{transprobdens}
  p_t(\bx,\by) \equiv \left( 2 \pi t \right)^{- \, d/2}
              \exp \left( - \frac{\|\bx-\by\|^2}{2t} \right).
\end{equation}
The initial value problem for the classical heat equation,
\begin{equation}
  \partial_t \, \varphi(t,\bx) = \frac{1}{2} \Delta \, \varphi(t,\bx), \qquad \varphi(0,\bx) = u(\bx), 
\end{equation}
is solved using the Brownian semigroup operators $P_t$,
\begin{equation}
  \varphi(t,\bx) = (P_t u) (\bx) := 
  \int d^dz \, u(\bz) p_t(\bx,\bz) \label{sg},
\end{equation}
i.e., as a linear superposition of the fundamental solutions $p_t$.

The semigroup property of the operator $P_t$,
\begin{equation}\label{semigroup}
  P_t \circ P_s = P_{t+s},
\end{equation}
implies the so-called Markov property of Brownian motion, i.e., any
point on the Brownian path can be considered as the starting point of
a new Brownian motion. In terms of the transition density $p_t$, the
Markov property hence takes the form 
\begin{equation}\label{Markov}
  p_{t+s}(\bx,\by) = \int d^dz\, p_t(\bx,\bz) p_s(\bz,\by).
\end{equation}
By definition of the expectation, we can rewrite (\ref{sg}) as
an integral over the space of continuous functions with respect to
the Wiener measure $\Pb_x$,
\begin{equation}
  \varphi(t,\bx) = \E_x \{ u(B_t) \} =
  \int_{C_0[0,\infty)} \Pb_x(d\omega) \, u(\omega_t). \label{pi}
\end{equation}
Hence, the solution of the initial value problem for the classical
heat equation can be represented either
using the Brownian semigroup or, equivalently, as a path integral.
An important characteristic of Brownian paths was found by 
Paley, Wiener and Zygmund \cite{Paley}, who showed that with respect to the Wiener measure, 
almost all Brownian paths are nowhere differentiable.  We will come
back to this point in section \ref{sec_singlesystem}.\\
We emphasize that the heat equation (\ref{heat}) describes a
\textsl{single} Brownian particle, by the heat function $\varphi(t,\bx)$. 
As in the case of the Schr\"odinger equation, the formalism, as it stands, cannot be extended to 
accommodate a non-constant particle number, as this would require a
non-constant number of functions $\varphi_i$ in the dynamical equations.
 However, the desire to study killing
and branching processes has led to the following two constructions.\\

\subsection{Softly killed Brownian motion}
It is interesting to study the heat equation with a potential term $v(\bx)$, which is assumed to be bounded from below,
\begin{equation}\label{softkilled}
  \partial_t \varphi(t,\bx) = \frac{1}{2} \Delta \varphi(t,\bx) - v(\bx) \varphi(t,\bx).
\end{equation}
Given the initial condition $\varphi(0,\bx)=u(\bx)$, the solution to
(\ref{softkilled}) is given by the Feynman-Kac formula \cite{Karatzas}
\begin{equation}\label{feynmankac}
     \varphi(t,\bx) = \E_{\bx} \left\{ u(B_t) 
     \exp \left(- \int_0^t ds \, v(B_s) \right) \right\}.
\end{equation}
For a constant potential $v\equiv \gamma$, the solutions $\varphi(t,\bx)$ can be interpreted as the transition densities 
for a \textsl{softly killed} Brownian particle, surviving a time interval $\Delta t$ only with probability 
$\exp(- \gamma \Delta t)$. The terminology refers to the fact that we are dealing with an inherent one-particle 
formalism, and killing is a matter of interpretation of the underlying
random variables. This is seen most clearly when considering the continuous paths in Wiener space, 
which do not stop at a finite parameter value (`hard killing'), but must, in this
context, rather be understood as random elements in 
$C([0,\infty)) \times \reals^+$, i.e., as paths with a killing time 
assigned to them.

This extension of the Wiener space as the underlying probability space
is necessary in order to encode a process with non-constant particle
number, in a setting that \textsl{by construction} can only deal with
a constant number of particles. 

\subsection{Branching Brownian motion}\label{bndclass}       
The standard approach to branching Brownian motion \cite{Etheridge} is the study of trees composed of independent, softly 
killed one-particle Brownian motions, where a dying particle gives
birth to $k$ children, according to a probability distribution $(p_k)$
on $\{0, 1, 2, \dots\}$.
This is formalized by consideration of a measure-valued process
$\{X_t: t \geq 0\}$, defined by the counting 
measure
\begin{equation}
  X_t(C) := \#\{\textrm{particles in } C \textrm{ at time } t\},
\end{equation} 
where $C$ is a Borel set in $\reals^d$. Hence, if we represent each
individual particle at a point 
$\bx \in \mathbb{R}^d$ by
a unit point mass $\delta_\bx$ and write $Y^i_t$ for the position of the $i$-th particle alive at time $t$,
we have
\begin{equation}
  X_t = \sum_i \delta_{Y^i_t}.
\end{equation}
As the exponential distribution has no memory, the measure-valued process $X$ inherits the {\em Markov property}
from the underlying Brownian motion. 
Due to Mc Kean (1975), the distribution of branching Brownian motion can be characterized in terms of a
non-linear extension of the heat equation: Let $\psi$ be a positive, bounded and continuous function on $\mathbb{R}^d$ such that
$0\leq\phi (\bx)\leq1$ for all $x \in \mathbb{R}^d$.
Then the quantity
\begin{equation}
  \varphi(t,\bx) \equiv \E_{\delta_\bx} \Big\{ \prod_i \phi(Y^i_t) \Big\} 
\end{equation}  
satisfies the evolution equation
\begin{equation}\label{branching}
  \begin{array}{c}
  \partial_t \varphi = \frac{1}{2} \Delta \varphi +  \gamma (W(\varphi) - \varphi) \\
  \varphi(0,\bx) = \phi(\bx)
  \end{array},
\end{equation}
where $W(\varphi):=\sum_{k=0}^\infty p_k \varphi^k$ is the probability generating
function of the distribution $(p_k)$, controlling the fertility of a
dying particle. 
Now $u$ is interpreted as describing an average mass flow of the
measure-valued process $X$.
This looks like a straightforward extension of the original heat equation by a non-linear term. However, 
the paths are now measure-valued, and hence we again deal with a somewhat different mathematical object, 
despite the formal similarity of equations (\ref{heat}) and (\ref{branching}). 
These adaptations were necessary to construct a spatially branching model within a formalism that originally 
described a one-particle system.\\

One can extract the extinction probability for such a branching and
dying process via the following construction.
Denote the total mass process of the
branching Brownian motion $X$ by $N_t:=X_t(\mathbb{R}^d)$. It is well-known (see e.g.
\cite{Athreya}) that $\{N_t : t \geq 0\}$ is a continuous time
Galton-Watson branching process whose moment generating function is
given by
\begin{equation}\label{phitilde}
  \tilde\varphi_\theta(t) \equiv \mathbb{E} \{ \theta^{N_t} \,|\, N_0=1 \} =
  \sum_{k=0}^\infty \theta^k \, \mathbb{P} \{ N_t=k \,|\,N_0=1\},
\end{equation}
for $\theta \geq 0$ (and $0^0 \equiv 1$). The function
$\tilde\varphi_\theta$ solves the initial value problem
\begin{eqnarray}
  \partial_t \tilde\varphi_\theta(t) &=& \gamma \left(W(\tilde\varphi_\theta(t))-\tilde\varphi_\theta(t)\right), \label{dualN} \\
             \tilde\varphi_\theta(0) &=& \theta. \nonumber
\end{eqnarray}
Compared to equation (\ref{branching}), this evolution equation is
missing the Laplace operator that generates the spatial evolution of
the process. Hence,  equation (\ref{dualN}) describes the pure
branching process, without the spatial evolution. For the special case $\theta=0$ we have by (\ref{phitilde}) that
$\tilde\varphi_0(t)=\mathbb{P} \{N_t =0 \,|\, N_0=1\}$. The solution
of the initial value problem (\ref{dualN}), with $\theta=0$, hence is
the extinction
probability for the total mass process $N_t$ in the limit $t \rightarrow
\infty$. In terms of the process
described by (\ref{branching}), this corresponds to the probability
that the process dies out eventually. We will
find equation (\ref{dualN}, $\theta\equiv 0$) as the differential form
of the recurrence relation for the so-called one-point function in our
field-theoretical treatment in Section \ref{sec_bnd}.

\subsection{Discussion of the traditional approach}\label{sec_discuss}
The one-particle Brownian motion is mathematically well-understood, and allows for a rigorous probability 
theoretical definition in terms of Wiener spaces. The desire to study
more complex Markov processes, at the same level of mathematical rigor,
led to the 
introduction of an exponential clock and the consideration of measure-valued paths. These generalizations of 
the one-particle case, however, come at the cost of an extension of
the underlying probability spaces, generalizing paths $C([0,\infty))$
to $C([0,\infty)) \times \reals^+$, or even measure-valued paths, as
explained in the preceding subsections. 
The necessity to perform these rather complicated and technical adaptations roots in the fact that these more complex Markov processes all 
feature a \textsl{non-constant particle number}, which must be encoded in a one-particle formalism, in a manner that varies from case to case.\\
The problem of dealing with systems of non-constant particle number is not a new one, but appears 
naturally in relativistic quantum physics. There, the relativistic mass-energy equivalence $E=mc^2$ allows for particle 
creation and annihilation, and hence is a feature of any quantum particle process at sufficiently high energy. 
It is clear that the Schr\"odinger equation, 
\begin{equation}\label{schroed}
  i \partial_t \psi(t,x) = - \frac{1}{2} \Delta \psi(t,x) + V(x) \psi(t,x),
\end{equation}  
describing the quantum behaviour of a one-particle quantum system in a
potential $V$, cannot feature such a dynamical creation 
and annihilation of particles, due to the fixed number of functions $\psi$. An adequate solution was found in form
of quantum field theories \cite{Peskin}. It was recognized \cite{Nagasawa} that
quantum field theory can be formally applied to the
discussion of Brownian motion, and vice versa, if one replaces the
time $t$ by imaginary time $it$. This ad-hoc adaptation, however,
bypasses the whole construction of a canonically quantized theory
starting from its classical equations. There are \textsl{no} insights gained as
to why one has to consider imaginary time, what the underlying
geometrical structures are, or why there is no interference of heat in
such a treatment. In fact, there exists no well-defined measure \cite{Keller}
for a Feynman path integral
which formally solves the Schr\"odinger equation (\ref{schroed}). 
Therefore, it seems that using quantum field theoretical methods
to describe Brownian processes might imply a loss in mathematical
rigour. In this paper, however, we show that careful consideration of the
structure of the heat phase space leads to a canonical quantization of
heat, whose path integral is well-defined.\\
In order to understand the occurrence of the imaginary unit in the
quantum mechanical
'path integral', it is instructive to re-write the
Lagrangian density for the free ($v=0$) Schr\"odinger equation (\ref{schroed}),
\begin{equation}
  \mathcal{L} = \frac{i}{2}(\partial_t\psi \psi^* - \partial_t\psi^* \psi) -
  \frac{1}{2}\nabla\psi\nabla\psi^*,
\end{equation}
in terms of the real and imaginary parts $\alpha$ and $\beta$ of the complex-valued probability amplitude $\psi = \alpha + i \beta$,
\begin{equation}
  \mathcal{L} = (\partial_t\alpha\beta - \alpha\partial_t\beta) - \frac{1}{2}(\nabla\alpha\nabla\alpha + \nabla\beta\nabla\beta). 
\end{equation}  
The imaginary part $\beta$ is recognized as the canonical momentum associated with the real part $\alpha$,
\begin{equation}
  \pi_\alpha \equiv \frac{\partial \mathcal{L}}{\partial \partial_t\alpha} = \beta,
\end{equation}
and the time evolution of the field phase space point $(\alpha,\beta)$
thus reads 
\begin{equation}
  \frac{d}{dt}\left[\begin{array}{c}\alpha\\\beta\end{array}\right] =
  \left[\begin{array}{cc}0 & -1\\1&0\end{array}\right] \, \frac{1}{2}
  \Delta \, \left[\begin{array}{c}\alpha\\\beta\end{array}\right],
\end{equation}
revealing the complex structure of the field phase space of the
Schr\"odinger equation. This complex structure is the actual reason
for the imaginary unit entering the exponential in the quantum
mechanical analogue of the Feyman-Kac path integral
(\ref{feynmankac}). It is therefore worthwhile to investigate the phase
space structure of the heat equation in the following section.\\

\section{Heat phase space}
This chapter studies the geometry of the heat phase space. The
central result is the absorption of the occurring structure into the
ring of pseudo-complex numbers. For the discussion of Brownian motion,
these play as important a r\^ole as the complex numbers do in quantum
mechanics.

\subsection{The structure of heat phase space}
In order to identify the canonical momentum associated with the heat
$\phi$, we need to devise a Lagrangian density for the heat
equation. This, however, can only be done by introduction of a second
dynamical field $\eta$. Without such, any attempt to generate the
first order time derivative in (\ref{heat}) from a Lagrangian density
inevitably results in a total time derivative, and hence does not
contribute to the Euler-Lagrange equations at all. We therefore
consider the Lagrangian density 
\begin{equation}\label{heatampllag}
  \mathcal{L} = (\partial_t\varphi \eta - \varphi \partial_t \eta) + \nabla \varphi \nabla \eta,
\end{equation}
yielding the Euler-Lagrange equations
\begin{eqnarray}
  \partial_t\varphi &=& \quad \frac{1}{2} \Delta \varphi\\
  \partial_t\eta &=& -\frac{1}{2} \Delta \eta.
\end{eqnarray}
We observe that the additional dynamical field $\eta$, required by a Lagrangian treatment of heat, is the canonical momentum density associated with the heat field $\varphi$, as
\begin{equation}
  \pi_\varphi \equiv \frac{\partial\mathcal{L}}{\partial \partial_t\varphi} = \eta.
\end{equation}  
The phase space picture of the heat evolution,
\begin{equation}\label{heatamp}
  \frac{d}{dt}\left[\begin{array}{c}\varphi\\\eta\end{array}\right] =
  \left[\begin{array}{cc}1 & 0\\0&-1\end{array}\right] \, \frac{1}{2}
  \Delta \, \left[\begin{array}{c}\varphi\\\eta\end{array}\right],
\end{equation}  
reveals that the geometry of the heat phase space is governed by a
\textsl{product structure} \cite{Yano}, i.e. the representing matrix squares to
the identity.
The above matrix representation of the product structure and the phase space element $\phi \equiv [\varphi,\eta]$ clearly depend on a particular choice of basis in phase space. In the following, we will show that there is a basis-independent formulation
\begin{equation}\label{heatamplitude}
  I \partial_t \phi = \frac{1}{2} \Delta \phi 
\end{equation}   
for equation (\ref{heatamp}). The obvious similarity of (\ref{heatamplitude}) with the Schr\"odinger equation (\ref{schroed}) suggests to term $\phi$ the \textsl{heat amplitude}, and equation (\ref{heatamplitude}) the \textsl{heat amplitude equation}. 

Consider a linear operator $I$ on $\reals^2$, with the
properties $I^2 = \textrm{id}_{\reals^2}$ and $I \neq
\pm \textrm{id}_{\reals^2}$. Then $\sigma_\pm \equiv
\frac{1}{2}(\textrm{id} \pm I)$ are orthogonal projection operators,
i.e., $\sigma_\pm^2 = \sigma_\pm$ and $\sigma_+ \sigma_- = 0$. As
$\{\sigma_+, \sigma_-\}$ are linearly independent over $\reals$, they
provide a basis for $\reals^2$. Thus the operator identity
\begin{equation}
  I \sigma_\pm = \pm \sigma_\pm
\end{equation}
can be read as an eigenvalue equation for the operator $I$, regarding $\sigma_\pm$ as \textsl{vectors} of $\reals^2$.
This shows that the matrix representation in (\ref{heatamp})
corresponds to the choice of the $\sigma_\pm$-basis for (\ref{heatamplitude}). In particular, the heat amplitude is given by the $\reals^2$-valued function
\begin{equation}
  \phi = \varphi \sigma_+ + \eta \sigma_-.
\end{equation}
From this point of view, the abstract reasoning started after (\ref{heatamp}) simply
reproduced the previous results. Now we note, however, that there is a
\textsl{canonical multiplication} defined on the heat phase space
$\reals^2$, induced by the composition of operators
$\sigma_\pm$. Hence, the heat phase space is not merely a vector space, but a commutative ring! We pause in our development of the physical aspects, in order to explore this structure a bit further.\\

\subsection{The pseudo-complex ring}\label{sec_pring}

Let $F$ be the field of real numbers $\reals$ or complex numbers
$\mathbb{C}$. The pseudo-complex ring over $F$ \cite{FPS} is the set
\begin{equation}
  \pc \equiv \{p = a+ I b | a, b \in F\},
\end{equation}
equipped with addition and multiplication laws induced by those on
$F$, with the additional rule that $I \not\in F$ is a product
structure, i.e. $I^2=1$. It is easily verified that $\pc$ is a
commutative unit ring with zero divisors $\pc^0 \equiv \pc^0_+ \cup
\pc^0_-$, where
\begin{equation}
  \pc^0_\pm \equiv \langle\sigma_\pm\rangle_F.
\end{equation}
It is often convenient to employ the zero-divisor representation of a
number $p\in\pc$,
\begin{equation}
  p = a + I b = (a+b)\sigma_+ + (a-b)\sigma_-.
\end{equation} 
The zero-divisor branches $\pc^0_+$ and $\pc^0_-$ are the only multiplicative ideals in $\pc$, thus they are both maximal ideals. Hence, the linear maps
\begin{eqnarray}
  \Gamma_\pm: \pc &\longrightarrow& \pc/\pc^0_\pm \cong \pc^0_\mp \cong F,\\
  \Gamma_\pm(a+Ib) &=& a \pm b,
\end{eqnarray}
are the only non-trivial ring homomorphisms from $\pc$ to $F$,
i.e., for all $p, q \in \pc$
\begin{eqnarray}
  \Gamma(p + q) &=& \Gamma(p) + \Gamma(q),\\
  \Gamma(p q) &=& \Gamma(p)\Gamma(q).
\end{eqnarray}
Application of $\Gamma_\pm$ to the heat amplitude $\phi$ extracts the
heat $\varphi$ and antiheat $\eta$, respectively:
\begin{eqnarray}
  \Gamma_+(\phi) &=& \varphi,\label{extract}\\
  \Gamma_-(\phi) &=& \eta. 
\end{eqnarray} 
Physically, this means that heat amplitudes do not interfere
(linearity), and that one can compose heat amplitudes for complicated
processes multiplicatively from elementary ones. In other words, the
extraction of heat from heat amplitudes is compatible with the Markov
property (\ref{semigroup}) of Brownian motion.
Finally, we define the linear map $*: \pc \longrightarrow \pc$ by its
action $* \sigma_\pm \equiv \sigma_\mp$. Clearly, $*$ is an
involution, as $**=1$. Acting on the heat amplitude, $*$ effects an
exchange of heat and antiheat. 


We have seen, from the analysis of the phase space structure of the
heat equation, that the heat amplitude is a pseudo-complex valued
field. For the discussion of Brownian motion, pseudo-complex numbers
therefore play the same r\^ole as complex numbers do in quantum
mechanics. The systematic exploitation of this insight will be
seen to be intimately linked to the intrinsical properties of Brownian
motion, like the non-differentiability of Brownian paths.

\subsection{Stationary solutions and unitary time evolution}
It is often sufficient to restrict the discussion of solutions $\phi(x)$ of the heat amplitude equation (\ref{heatamplitude}) to the solutions $\Phi(\bx)$ of the \textsl{stationary} heat amplitude equation
\begin{equation}\label{stateqn}
  \frac{1}{2} \Delta \Phi(\bx) = - E \Phi(\bx), \qquad E \in \reals^+,
\end{equation} 
which can be extended to solutions of the time-dependent dynamics by acting on them with the time evolution operators
\begin{equation}\label{timeevolop}
  U(t) \equiv \exp(- I E t),
\end{equation}
such that $\phi(t,\bx) = U(t) \Phi(\bx)$ solves (\ref{heatamplitude}). Note that the time evolution operators form a semigroup, i.e.,
\begin{equation}
  U(t) \circ U(s) = U(t+s),
\end{equation} 
and are unitary with respect to pseudo-complex conjugation,
\begin{equation}
 U^*(t) = U^{-1}(t) = U(-t). 
\end{equation}
In section \ref{sec_singlesystem} we will identify the stationary
solutions $\Phi$ as elements of a Hilbert module $\mathcal{H}$, and the time
evolution operators (\ref{timeevolop}) as unitary operators on $\mathcal{H}$. 

\subsection{Lagrangian for the heat amplitude equation}
Equipped with the constructions of the preceding sections, 
we are now in a position to write down the Lagrangian density (\ref{heatampllag}) in terms of the pseudo-complex valued heat amplitude, i.e., 
\begin{equation}
  \mathcal{L} = \frac{I}{2} (\phi\partial_t\phi^* - \partial_t\phi\phi^*) - \frac{1}{2} \nabla \phi \nabla \phi^*.
\end{equation} 
Note that the Lagrangian density is real-valued. For later developments, it will prove useful to consider a slightly more
general form, namely including a
quadratic term
\begin{equation}\label{clockmodel}
  \mathcal{L}_B =  \frac{1}{2} I ( \phi \partial_t \phi^* - \phi^* \partial_t \phi
  ) - \frac{1}{2}\nabla \phi \nabla \phi^* - \gamma \phi \phi^*,  
\end{equation}
which will be seen to give rise to an exponential clock with clock
rate $\gamma$. Observe that the Euler-Lagrange equations for
(\ref{clockmodel}) yield the equation of motion (\ref{softkilled}) for
softly killed Brownian motion with a constant potential $v(x) \equiv \gamma$.
Then the canonically conjugate fields to $\phi, \phi^*$ are
\begin{eqnarray}
  \pi \equiv \frac{\partial\mathcal{L}}{\partial\partial_t\phi} &=& -
  \frac{I}{2}\phi^*,\label{pifield}\\
  \pi^* \equiv \frac{\partial\mathcal{L}}{\partial\partial_t\phi^*} &=& 
  \frac{I}{2}\phi,
\end{eqnarray}
and we obtain the Hamiltonian density $\mathcal{H}$ as the Legendre transform of $\mathcal{L}$,  
\begin{equation}
  \mathcal{H} \equiv \pi \partial_t \phi + \pi^* \partial_t \phi^* -
  \mathcal{L} =  \frac{1}{2} \nabla \phi \nabla \phi^* + \gamma \phi \phi^*.
\end{equation}
We observe that the Hamiltonian density is real-valued, such that the Hamiltonian 
\begin{equation}
  H \equiv \int d^dx \mathcal{H}
\end{equation}
can be consistently interpreted as the energy of the heat amplitude field. 

\section{Single particle system}\label{sec_singlesystem}
For the intended development of the multi-particle theory in section
\ref{sec_multi}, a thorough understanding of the one-particle system
is inevitable. It is convenient to introduce a basis free Dirac notation for the heat
amplitude equation (\ref{heatamplitude}), in terms of a Hilbert module
formalism. 

\subsection{Hilbert modules over $\pc$}
Let $(\mathcal{R}, \bracket{\cdot}{\cdot}_\reals)$ be a Hilbert vector
space over $\reals$. We define the associated \textsl{Hilbert $\pc$-module}
$(\mathcal{H},\bracket{\cdot}{\cdot})$ by
\begin{equation}
  \mathcal{H} = \mathcal{R} \sigma_+ \oplus \mathcal{R} \sigma_-,
\end{equation}
with an \textsl{indefinite} inner product $\bracket{\cdot}{\cdot}$ on
$\mathcal{H}$ defined by
\begin{equation}
  \bracket{f}{g} \equiv \bracket{f_-}{g_+}_\reals \sigma_+ + \bracket{f_+}{g_-}_\reals \sigma_-.
\end{equation}
It follows immediately that $\bracket{\cdot}{\cdot}$ is pseudo-hermitian and semi-linear in the first argument,
\begin{eqnarray}
  \bracket{f}{g} &=& \bracket{g}{f}^*,\\
  \bracket{\lambda f}{\mu g} &=& \lambda^* \mu \bracket{f}{g},\qquad
  \lambda, \mu \in \pc.
\end{eqnarray}
As any real Hilbert space $\mathcal{R}$ has a basis, so has
$\mathcal{H}$, i.e., $\mathcal{H}$ is a free module. 
An operator $A = A_+ \sigma_+ + A_- \sigma_-$ on $\mathcal{H}$ is seen
to be self-adjoint with
respect to $\bracket{\cdot}{\cdot}$ if, and only if, its zero-divisor components
$A_\pm: \mathcal{R} \longrightarrow \mathcal{R}$ satisfy
\begin{equation}
  A_+ = A_-^\#,
\end{equation}
where $\#$ denotes the adjoint with respect to the inner product
$\bracket{\cdot}{\cdot}_\reals$ on $\mathcal{R}$. We further call an
operator $A: \mathcal{H} \longrightarrow \mathcal{H}$ \textsl{real} if
$A=A^*$. It follows immediately that a self-adjoint real operator on
$\mathcal{H}$ is a self-adjoint operator on $\mathcal{R}$. The
spectral theorem for the Hilbert vector space $\mathcal{R}$ therefore guarantees that the Hilbert module
$\mathcal{H}$ decomposes into a direct sum of pseudo-complex eigenspaces $E_\lambda$
of any (i) selfadjoint and (ii) real operator $A$ on $\mathcal{H}$,
\begin{equation}
  \mathcal{H} = \bigoplus_\lambda E_\lambda(A).
\end{equation}
Note that this slightly extended form of the spectral theorem for
Hilbert modules over $\pc$ does not recur to the underlying real
Hilbert space $(\mathcal{R},\bracket{\cdot}{\cdot}_\reals)$ any more,
but is entirely formulated in terms of the Hilbert module $(\mathcal{H},\bracket{\cdot}{\cdot})$.
An important example of a Hilbert module is the space of square-integrable pseudo-complex
valued functions on $\reals^d$, with the indefinite inner
product $\bracket{f}{g} \equiv \int d^dx f^*(\bx) g(\bx)$. The
operator $\mathbf{\hat X} \equiv \bx$ is self-adjoint and real, and
hence gives rise to a complete basis of $\mathcal{H}$. In contrast,
the operator $\mathbf{\hat P} \equiv - I \nabla_\bx$ is
only self-adjoint, but not real. Hence the spectral theorem does not
apply in the case of $\mathbf{\hat P}$. In the next section, we will
see that the spectrum of $\mathbf{\hat P}$ is indeed empty.

Now let $\{\ket{\bx}\}$ be a complete
orthonormal basis of eigenvectors of a self-adjoint real operator
$\mathbf{\hat X}: \mathcal{H} \longrightarrow \mathcal{H}$, i.e.,
\begin{eqnarray}
  \mathbf{\hat X} \ket{\bx} &=& \bx \ket{\bx},\\
  \bracket{\bx}{\by} &=& \delta(\bx-\by),\\
  \textrm{id}_\mathcal{H} &=& \int dx \ket{\bx}\bra{\bx}.
\end{eqnarray} 
We will refer to components $\bracket{\bx}{f}$ of general states $\ket{f} \in
\mathcal{H}$ with respect to the $\ket{\bx}$-basis as the \textsl{position
space representation} of $\ket{f}$. Stationary solutions (\ref{stateqn}) of the heat
amplitude equation are elements
$\ket{\Phi}\in\mathcal{H}$. Define states $\ket{\bp}$ by their components in the position space
representation,
\begin{equation}
  \bracket{\bx}{\bp} \equiv \cos(\bp\bx) + I \sin(\bp\bx),
\end{equation} and observe that we can, in turn, express the $\{\ket{\bx}\}$
in the $\ket{\bp}$ basis,
\begin{equation}
  \ket{\bx} = \int dp \ket{-\bp}\bracket{\bp}{\bx}.
\end{equation}
Hence $\{\ket{\bp}\}$ constitutes another complete basis of $\mathcal{H}$, which
one may call \textsl{twisted orthonormal}, because of the relation
\begin{equation}
  \bracket{\bp}{-\bq} = \delta(\bp-\bq).
\end{equation}
It is sometimes useful to insert the identity in the form of the
twisted orthonormal basis,
\begin{equation}
  \textrm{id}_{\mathcal{H}} = \int dp \ket{-\bp}\bra{\bp}.
\end{equation}

\subsection{Single particle Hilbert formalism}
We now put the abstract developments of the previous subsection to use
in the discussion of a single Brownian particle. Define an operator
$\mathbf{\hat P}$ by its position space
representation 
\begin{equation}
  \bra{\bx} \mathbf{\hat P} \ket{\by} \equiv - I\, \delta(\bx-\by)\, \nabla_\bx,\end{equation} 
and note its self-adjointness $\mathbf{\hat P}^+ = \mathbf{\hat P}$. The unitary
operator $\exp(I \mathbf{a}.\mathbf{\hat P})$, acting on the Hilbert
module, translates states in position space by $\mathbf{a}$. We
therefore identify its generator $\mathbf{\hat P}$ as the momentum
operator for the one-particle states.  
We observe that the states $\ket{\bp}$ satisfy the stationary heat
amplitude equation
\begin{equation}\label{stationaryDirac}
  \hat H \ket{\bp} = \half \bp^2 \ket{\bp},
\end{equation} with $\hat H \equiv \frac{1}{2} \mathbf{\hat P}^2$, such that
the full time-dependent solution $\ket{\bp}$ is given by
\begin{equation}\label{timeevol}
  \ket{\bp(t)} \equiv \exp(- {\frac{I}{2}}\bp^2 t) \ket{\mathbf{p}}
\end{equation}
Hence, any linear combination $\ket{\phi(t)} = \int dp \, \tilde\phi(\bp)
\ket{\bp(t)}$ satisfies the time-dependent heat amplitude equation
\begin{equation}
  I \partial_t \ket{\phi(t)} = \hat H \ket{\phi(t)}.
\end{equation}
We emphasize that the states $\ket{\bp}$ are \textsl{not} eigenstates of the
momentum operator $\hat P$, because we have
\begin{equation}
  \mathbf{\hat P} \ket{\bp} = - \bp \ket{-\bp}.
\end{equation}
This shows that the fundamental solutions of the heat amplitude
equation have no well-defined momentum, but a well-defined
energy. Because the spectrum of $- I \nabla_\bx$ lies in $iI\reals$,
there are also no linear combinations over $\pc_\reals$ with a
well-defined momentum. This corresponds to the classical result that Brownian paths
are almost surely nowhere differentiable.
The probability density (\ref{transprobdens}) for a Brownian particle
to propagate from $\bx$ to $\by$ within time $t$, is given in Dirac
notation as the overlap of the initial and final states:
\begin{equation}\label{oneparttrans}
  p_t(\bx,\by) = \Gamma_+ \bracket{\by(t)}{\bx(0)},
\end{equation}
after projection $\Gamma_+$ to the heat part, according to our
discussion of the heat projection operator $\Gamma_+$ in section \ref{sec_pring}.
Using that $\Gamma_+: \pc \longrightarrow \reals$ is a ring homomorphism, it is 
easy to show that the transition probabilities defined by (\ref{oneparttrans}) satisfy the Markov property (\ref{Markov}).
This concludes our discussion of one-particle Brownian motion in terms
of the heat amplitude equation.\\

\section{Multi-Particle systems}\label{sec_multi}
We now want to give the idea of a multi-particle system a rigorous
meaning. The basis for the understanding of a multi-particle system is
the understanding of the one-particle system given by a Hilbert module
$(\mathcal{H}, \bracket{\cdot}{\cdot})$ and an equation
of motion $I \partial_t
  \ket{\phi(t)} =\hat H \ket{\phi(t)}$, as discussed in the previous section.
In the one-particle theory, the state of a system is given as an
element of the space $\mathcal{H}$.

\subsection{Fock space}
The one-particle space $\mathcal{H}$ is now extended, in order to accommodate the case of zero, one,
two, ... particles. the appropriate structure is the so-called
\textsl{Fock space} generated by $\mathcal{H}$
\begin{equation}
  F(\mathcal{H}) \equiv \bigoplus_{n=0}^\infty \bigotimes_{i=1}^n
  \mathcal{H} \equiv \pc \oplus \mathcal{H} \oplus \mathcal{H} \otimes
  \mathcal{H} \oplus \dots, 
\end{equation}
where $\mathcal{H}^0 \equiv \pc$.
We will consider indistinguishable particles
obeying Bose statistics, i.e., an exchange of two particles goes
unnoticed. This choice is reflected in considering the boson Fock
space
\begin{equation}\label{bosonfock}
  \odot \mathcal H \equiv \bigoplus_{n=0}^\infty \bigodot_{i=1}^n \mathcal{H},
\end{equation}
where $\odot$ denotes a symmetrized tensor product. The symmetrization
removes any order of the factors in any of the spaces
$\mathcal{H}^{\odot n}$, which therefore encodes the
indistinguishability of the particles represented by these
factors. Antisymmetrization would also lead to indistinguishable
particles, but an exchange of an even number of them would not go
completely unnoticed, but rather effect a sign change.

The one-dimensional space
$\mathcal{H}^0$ accommodates a system of zero particles, and we call
its normalized basis vector
\begin{equation}
  \ket{\Omega} \equiv 1 \oplus 0 \oplus 0 \odot 0 \oplus \dots \in
  \odot \mathcal{H}
\end{equation}
the \textsl{vacuum} of the multi-particle theory. The vacuum is a new
concept that does not occur in the one-particle theory.
The boson Fock space $\odot \mathcal{H}$ inherits the inner product on
$\mathcal{H}$ in a natural way,
\begin{equation}\label{inducedprod}
  \bracket{u_1 \odot \dots \odot u_n}{v_1 \odot \dots \odot v_n}
  \equiv \bracket{u_1}{v_1} \dots \bracket{u_n}{v_n}.
\end{equation}
Define a set of linear operators $a^+_{\ket{\bp}}$ acting on the Fock space $\odot\mathcal{H}$, labelled by elements of
the one-particle Hilbert space $\mathcal{H}$. Their action on the
$n$-particle subspace $\mathcal{H}^{\odot n}\subset \odot\mathcal{H}$ is defined by
\begin{equation}\label{creationdef}
  a^+_{\ket{\bp}} \ket{u_1} \odot \dots \odot \ket{u_n} \equiv
  \ket{\bp} \odot \ket{u_1} \odot \dots \odot \ket{u_n}. 
\end{equation} 
In particular, we can generate the entire one-particle space
$\mathcal{H}$ from the vacuum,
\begin{equation}\label{onepartstate}
  a^+_{\ket{\bp}} \ket{\Omega} = \ket{\bp}.
\end{equation} 
The operators $a^+$ are called \textsl{creation operators}. Similarly,
we define linear \textsl{annihilation operators} $a^-_{\bra{\bp}}: \odot\mathcal{H} \longrightarrow \odot\mathcal{H}$, labelled
by elements of the dual space $\mathcal{H}^*$. They act on an element
of the $(n+1)$-particle subspace $\mathcal{H}^{\odot(n+1)}$ by
\begin{equation}\label{annihildef}
  a^-_{\bra{\bp}} \ket{u_0} \odot \dots \odot \ket{u_n} \equiv \sum_{i=0}^n
  \bracket{\bp}{u_i} \ket{u_0} \odot \dots \ket{\hat u_i} \odot \dots
  \odot \ket{u_n},
\end{equation}
where the hat on $\ket{\hat u_i}$ denotes the omission of this vector
in the tensor product.
In particular, all $a^-$ annihilate the vacuum,
\begin{equation}
  a^-_{\bra{\bp}} \ket{\Omega} = 0 \oplus 0 \oplus 0 \odot 0 \oplus \dots.
\end{equation}
Linearity allows to extend the definitions of $a^+, a^-$ to the entire
Fock space \cite{StochQuant}.
One easily checks that
\begin{equation}
  \left[a^-_{\bra{\bp}}, a^+_{\ket{\bq}}\right] = \bracket{\bp}{\bq}
\end{equation}
from definitions (\ref{creationdef}) and (\ref{annihildef}).
The choice of Bose statistics for the multi-particle spaces immediately
requires
\begin{eqnarray}
  \left[a^+_{\ket{\bp}}, a^+_{\ket{\bq}}\right] &=& 0,\\
  \left[a^-_{\bra{\bp}}, a^-_{\bra{\bq}}\right] &=& 0,
\end{eqnarray}
such that $a^+_{\ket{\bp}} a^+_{\ket{\bq}} \ket{\Omega} \equiv \ket{\bp} \odot \ket{\bq}$ is well-defined.
An important property for calculations is that creation and
annihilation operators are adjoint with respect to the inner product
(\ref{inducedprod}),
\begin{equation}
  \big<a^+_{\ket{\bp}} x \big | y\big> = \big<x \big| a^-_{\bra{\bp}} y\big>,
\end{equation} 
for $x\in \mathcal{H}^{\odot n}$, $y \in \mathcal{H}^{\odot (n+1)}$.
Finally, we define an operator-valued field
\begin{equation}
  \hat\Phi(\bx) \equiv \int \frac{d^dp}{(2\pi)^{d/2}} \,\, a^-_{\bra{\bp}} \bracket{\bx}{\bp},
\end{equation}
from which we can extract the annihilation operators by
\begin{equation}
  a^-_{\bra{\bp}} = (2\pi)^{d/2} \int dx \,\, \hat\Phi(\bx) \bracket{-\bp}{\bx}.
\end{equation}
Note that up to this point, the creation and annihilation operator gymnastics
have no physical meaning. They just describe the 'kinematics' of a
multi-particle system, independent of any particular dynamics. 

\subsection{Second quantization of heat}
We now
want to derive the stationary dynamics of the free multi-particle
theory from the stationary dynamics (\ref{stationaryDirac}) of the one-particle sector $\mathcal{H}$. To this end we
require that the one-particle states (\ref{onepartstate}) of the
multi-particle theory still satisfy the equation (\ref{stationaryDirac}),
\begin{equation}\label{hilfs}
  \hat H a^+_{\ket{\bp}} \ket{\Omega} = \frac{1}{2} \bp^2
  a^+_{\ket{\bp}} \ket{\Omega}.
\end{equation}
The same way in which we understand the one-particle states in terms of
creation and annihilation operators, we can construct the appropriate
'Hamiltonian' $\hat H$ in terms of creation and annihilation operators, 
\begin{equation}\label{multiham}
  \hat H \equiv \int d^dp \frac{\bp^2}{2} a^+_{\ket{-\bp}} a^-_{\bra{\bp}},
\end{equation}
such that (\ref{hilfs}) holds, as is easily checked using the
commutation relations for the $a^+, a^-$. Note the sign of the label
of the creation operator in (\ref{multiham}), due to the
pseudo-complex structure. 
It is now possible to express the Hamiltonian in terms of the operator-valued
field $\hat\Phi$,
\begin{equation}\label{phihamil}
  \hat H = \int d^dx \frac{1}{2} \nabla \hat\Phi(\bx) \nabla \hat\Phi^+(\bx),
\end{equation}
where $\Phi^+$ denotes the adjoint of $\Phi$ with respect to the inner
product on $\mathcal{H}$.

We know that the time-dependent one-particle states evolve
according to (\ref{timeevol}). In the multi-particle theory, it is
convenient to transfer this time dependence of the states to the operators
$a^+, a^-$ or, equivalently, $\hat\Phi$. Clearly, this change to the
so-called \textsl{Heisenberg picture} must preserve the inner product between
states. If we denote the time evolution operator $U(t) \equiv \exp(I\hat H t)$, and note its
unitarity $U U^+ = 1$ with respect to the inner product on $\mathcal{H}$,
then the Heisenberg picture $\phi(t,\bx)$ of the stationary field $\hat
\Phi(\bx)$ must be given by
\begin{equation}
  \hat\phi(x) \equiv \hat\phi(t,\bx) \equiv U(t) \hat\Phi(\bx) U^+(t),
\end{equation}
in order to preserve the inner product on $\mathcal{H}$. This
definition can be rephrased as the Heisenberg equation of motion
\begin{equation}
  I \partial_t \hat\phi(x) = \left[\hat H, \hat\phi(x)\right], 
\end{equation}
using that $H$ is self-adjoint.
Using the identity $[A,BC] = [A,B]C + B[A,C]$ and (\ref{phihamil}), one can
verify that the operator-valued field $\hat\phi(x)$ satisfies
\begin{equation}\label{opvalheat}
  I \partial_t \hat\phi(x) = \frac{1}{2} \Delta \hat\phi(x).
\end{equation}
This is the most important result of this section: We have shown that
the dynamics of the one-particle sector fully determine the dynamics
of the operator-valued field $\hat\phi(x)$, i.e. the dynamics of the
creation and annihilation operators! Indeed, the Lagrangian density for (\ref{opvalheat}),
\begin{equation}\label{oplag}
  \mathcal{L} = \frac{I}{2} (\phi^+ \partial_t \phi - \phi \partial_t 
\phi^+) - \half
\nabla \phi^+ \nabla \phi,
\end{equation}
gives rise to the Hamiltonian (\ref{phihamil}). Note that now the
ordering of the fields in (\ref{oplag}) is relevant, and must be
chosen as above to yield (\ref{phihamil}).  
An explicit integral representation for the Heisenberg picture field
$\hat \phi(x)$,
whose equations of motion follow from the more general Lagrangian density
(\ref{clockmodel}), including a quadratic term $\gamma \phi \phi^+$,
can be obtained by observing that
\begin{equation}
  \hat H^n a_\bp = a_\bp(\hat H - \frac{1}{2}\bp^2 - \gamma)^n,
\end{equation} 
which is immediate by an induction argument.
It follows that for these slightly more general dynamics, the
Heisenberg picture field $\hat \phi$ is given by the explicit expression
\begin{equation}\label{phiheisenberg}
  \hat \phi(x) = \int \frac{d^dp}{(2\pi)^{d/2}} a_\bp \exp(-I(\frac{1}{2}\bp^2+\gamma) x^0)
  \left\{\cos(\bp\bx) + I \sin(\bp\bx)\right\}.
\end{equation}
For completeness, we remark that the field $\hat\phi$ and its
canonically conjugate field $\hat\pi \equiv
\frac{\partial\mathcal{L}}{\partial \partial_t\hat\phi} = -\frac{I}{2} \hat\phi^+$ satisfy the \textsl{equal time} commutation relations
\begin{eqnarray}
  [\hat\phi(t,\bx), \hat\pi(t,\by)] &=& - \frac{I}{2} \bracket{\bx}{\by},\\
  {}[\hat\phi(t,\bx), \hat\phi(t,\by)] &=& 0,\\
  {}[\hat\pi(t,\bx), \hat\pi(t,\by)] &=& 0,
\end{eqnarray}
as is easily checked from the commutation relations of the
time-independent $a^+$ and $a^-$ operators. These equal time
commutation relations contain no dynamical information, but simply
encode the multi-particle kinematics. 
The next subsection deals with the explicit calculation of
the commutator of fields at different times, yielding dynamical
information on the system.

\subsection{Propagator for free Brownian motion}
We are now in a position to determine the one-particle propagator for Brownian
particles within the formalism of the multi-particle theory. The result
will present a check on our construction, as it must coincide with the
transition amplitude (\ref{oneparttrans}) obtained in the one-particle
theory. However, the calculation demonstrates the abstract concepts
developed in the previous section at work, and we therefore present it
in some detail.
In order to slim down the notation, we will omit the hat on the
operator-valued field $\hat\phi$ from now on. We are no longer dealing with
the classical heat amplitude $\phi$, so that there is no danger of confusion. 
Consider the amplitude for a particle to be created at spacetime point
$x$ and to be destroyed at $y$,
\begin{equation}
  B(x,y) \equiv \bra{\Omega} \phi(y) \phi^+(x) \ket{\Omega}.
\end{equation}
It is convenient to rewrite this in terms of a commutator
\begin{equation}
  B(x,y) \equiv \bra{\Omega}[\phi(x), \phi^+(y)]\ket{\Omega}
\end{equation}
for the Heisenberg field $\phi$. From the expansion
(\ref{phiheisenberg}) we obtain
\begin{equation}
  B(x,y) = \int \frac{d^dp}{(2\pi)^d} \exp(-I(\frac{1}{2}\bp^2+\gamma)\Delta
  t) \cos(\bp.\Delta\bx),
\end{equation}
where $\Delta t \equiv y^0 - x^0$ and $\Delta \bx \equiv \by - \bx$. 
We are interested in the transition amplitude for a Brownian particle,
and hence must project $B(x,y)$ by $\Gamma_+$, yielding
\begin{equation}
  \Gamma_+ B(x,y) = \int \frac{d^dp}{(2\pi)^d}
  \exp\left(-(\half \bp^2+\gamma) \Delta t \right) \exp(i \bp.\Delta\bx),
\end{equation}
where we have used the symmetry of the integral in $\bp$ to extend the
$\cos(\bp.\Delta\bx)$ to $\exp(\bp.\Delta\bx)$. Completing the square
and shifting of the integration variable $\bp \rightarrow \bp + i
\frac{\Delta \bx}{\Delta t}$ gives
\begin{equation}
  \Gamma_+ B(x-y) = e^{-\gamma \Delta t} e^{-\frac{(\Delta \bx)^2}{2\Delta t}}
  \int_{\reals^d} \frac{d^dp}{(2\pi)^d} \exp(- \frac{1}{2}\Delta t
  \bp^2).
\end{equation}
We define the \textsl{retarded propagator}
\begin{equation}\label{retprop}
  B_R(x,y) \equiv \theta(y^0-x^0) B(x,y),   
\end{equation}
with $\theta(0)=0$, and extract the transition probability from the retarded propagator
via the projection (\ref{extract}),
\begin{eqnarray}
  p(x,y) &\equiv& \Gamma_+(B_R(x,y))\nonumber\\
         &=& \exp(-\gamma \Delta t) (2\pi \Delta t)^{-\frac{d}{2}} \exp\left(-\frac{(\Delta
         \bx)^2}{2\Delta t}\right). 
\end{eqnarray}
This is exactly the transition density for Brownian motion
in $d$ spatial dimensions for $\gamma=0$, already obtained in the
one-particle formalism. For $\gamma>0$, the Brownian
particle is seen to carry an independent exponential clock with clock rate
$\gamma$. Later, we will put this observation to use in the construction of  
branching Brownian motion, which possesses the Markov property 
if, and only if, the lifetime of each generic path is exponentially distributed (see section \ref{sec_bnd}).\\
Also observe that the multiplicativity of $\Gamma_+$ corresponds to the semigroup structure of
Brownian transition operators.
The \textsl{Feynman propagator} is defined as the \textsl{time-ordered} product
of field operators,
\begin{equation}\label{feynprop}
  B_F(x,y) \equiv \bra{\Omega} T\left\{\phi(x)\phi^+(y)\right\}
  \ket{\Omega}. 
\end{equation}
Using the integral representation 
\begin{equation}
  \theta(x^0) = \int \frac{dp^0}{2\pi} \frac{\sin(p^0x^0)}{p^0},
\end{equation}
one finds 
\begin{equation}
B_F(x,y) =  \int_{\reals^d}
\frac{d^{d+1}p}{(2\pi)^{d+1}}
\frac{1}{ip^0-\frac{\bp^2}{2}-\gamma} \exp\left\{-Ii k (y-x)\right\},
\end{equation}
where $k\equiv(k^0,\bk)$ and
the inner product is euclidean.\\

\section{Interacting Multi-Particle Systems}
Our criticism in section \ref{sec_discuss}, of the standard
treatment of Brownian processes, was focused on the fact that
such models are not fully specified by their dynamics, but also
require an undesirable case-to-case adaptation of the underlying
mathematical structures. The pseudo-complex Fock space formalism, developed in section
\ref{sec_multi}, provides the solution to this problem. One can now
study almost arbitrary extensions of the operator-valued field equation
(\ref{opvalheat}), in order to describe multi-particle systems, where
the particle number changes dynamically. In the field theoretical
framework, such dynamical creation and annihilation of particles are
already fully encoded in so-called interaction terms added to the free
Lagrangian. In this section, we first develop the theory a little for general
interaction terms $\mathcal{L}_{\textrm{{\small int}}}$. As an example, we
will study the binary branching and dying model in the next section.\\

\subsection{Unitary time evolution}
Consider Brownian motion with a general interaction term
$\mathcal{L}_{\textrm{\small int}}$, giving rise to a total Lagrangian
density
\begin{equation}  
  \mathcal{L} = \mathcal{L}_B + \mathcal{L}_{\textrm{int}}[\phi,\phi^*].
\end{equation}
In terms of the Hamiltonian formalism, this gives rise to an additional
interaction term 
\begin{equation}
  H_{\textrm{int}} = - \int d^dx \mathcal{L}_{\textrm{int}}
\end{equation}
in the complete Hamiltonian $H = H_0 + H_{\textrm{int}}$. It is
convenient to define the so-called \textsl{interaction picture} field
$\phi_I$ as the Heisenberg picture field of the free
theory,
\begin{equation}\label{interactfield}
  \phi_I(x) = \exp(I H_0 t) \Phi(\bx) \exp(-I H_0 t), 
\end{equation}
where $x \equiv (T,\bx)$
Hence, the full Heisenberg field of the interacting theory can be
expressed in terms of $\phi_I$ as
\begin{equation}\label{heisfield}
  \phi(x) = V(t)^+ \phi_I(x) V(t)
\end{equation}
with 
\begin{equation}
  V(t) = \exp(I H_0 t) \exp(- I H t).
\end{equation}
The unitary evolution operator $V(t)$ satisfies the differential equation
\begin{equation}\label{pseudoschroedinger}
  - I \partial_t V(t) = H_I(t) V(t),
\end{equation}
where 
\begin{equation}
  H_I(t) \equiv \exp(I H_0 t) H_{\textrm{int}} \exp(-I H_0 t) 
\end{equation}
is the interaction Hamiltonian in the interaction picture.
Equation (\ref{pseudoschroedinger}), with the initial condition $V(0)=1$ has the iterative solution 
\begin{equation}
  V(t) = 1 - I \int_0^t dt_1 H_I(t_1) V(t). 
\end{equation}
By iteration one gets, with $t_0 \equiv t$, 
\begin{equation}
  V(t) = \sum_{n=0}^\infty (-I)^n \prod_{i=1}^n
  \left(\int_0^{t_{i-1}} dt_i H_I(t_i)\right), 
\end{equation}
such that the introduction of a time ordering operator and appropriate
combinatorial factors \cite{Peskin} yields
\begin{eqnarray}
  V(t) &=& \sum_{n=0}^\infty \frac{(-I)^n}{n!} T
  \left(\int_0^{t_0} dt' H_I(t')\right)^n \nonumber\\
       &\equiv& T \exp\left(-I\int_0^{t_0} dt' H_I(t')\right).\label{interV} 
\end{eqnarray}

\subsection{Correlation functions}
The field $\phi$ encodes the full time-dependence of the
interacting theory, whereas $\phi_I$ simply describes the free theory,
according to its definition (\ref{interactfield}). The $(n,m)$-point correlation
function of the free theory is defined by
\begin{equation}
  \bra{\Omega}T\{\phi_I(x_1) \dots \phi_I(x_n) \phi_I^+(x_{n+1}) \dots \phi_I^+(x_{n+m}) \}\ket{\Omega},
\end{equation}
such that the Feynman propagator (\ref{feynprop}) is recognized as the
free two-point correlation function. The vacuum
$\ket{\Omega}$ is the ground state of the free theory, i.e., it is the
eigenstate of $H_0$ with minimal eigenvalue. Correspondingly, we define the
vacuum of the interacting theory, denoted by $\ket{\Sigma}$, as the
lowest eigenvalue eigenstate of the full Hamiltonian $H$. Using (\ref{interV}), it
is a standard exercise \cite{Peskin} to show that the $(n,m)$-point
correlation function for the interacting theory,
\begin{equation}
  \bra{\Sigma}T\{\phi(x_1) \dots \phi(x_n) \phi^+(x_{n+1}) \dots
  \phi^+(x_{n+m}) \} \ket{\Sigma},
\end{equation}
where $\phi$ is the Heisenberg picture field
(\ref{heisfield}) of the interacting theory,
can be calculated from the interaction picture field $\phi_I$ (\ref{interactfield}) and the
free vacuum $\ket{\Omega}$ as
\begin{equation}\label{corrfunc}
    \frac{
       \bra{\Omega} T\left\{ \phi_I(x_1) \dots \phi_I(x_n) \phi_I^+(x_{n+1}) \dots \phi_I^+(x_{n+m}) \exp(-I\int
          d^{d+1}z \mathcal{H}_I(z)) \right\} \ket{\Omega}
         }
         {
          \bra{\Omega} T \exp(-I\int
          d^{d+1}z \, \mathcal{H}_I(z))
          \ket{\Omega} 
         },
\end{equation}
where $\mathcal{H}_I$ is the Hamiltonian density for the interaction
Hamiltonian $H_I$. 
It is desirable to simplify this expression by replacing the
calculationally awkward time ordering operator $T$ by so-called
normal-ordered expressions. The following subsection
introduces Feynman diagrams as the appropriate technology.  

\subsection{Wick's Theorem and Feynman diagrams}
We introduce the normal ordering operator $N$, which acts on a product of creation and annihilation operators by regrouping the factors such that all annihilation operators
are sent to the right of all creation operators, e.g,
\begin{equation}
  N(a^+_{\ket{\bp}} a^-_{\bra{\bq}} a^+_{\ket{\mathbf{r}}}) \equiv
  a^+_{\ket{\bp}}  a^+_{\ket{\mathbf{r}}} a^-_{\bra{\bq}}.
\end{equation}
Wick's theorem states how a time-ordered product of heat amplitude fields $\phi_I$, as
occurring in (\ref{corrfunc}), can be expressed by normal ordered
expressions and contractions,
\begin{eqnarray}
  & &    T\left\{\phi_I(x_1) \phi_I(x_2) \phi_I^+(x_3) \dots \phi_I(x_f)\right\}\nonumber\\
  &=&  N \left[\phi_I(x_1) \phi_I(x_2) \phi_I^+(x_3) \dots
  \phi_I(x_f)\right]\\
  & &  + N \left[\textrm{all (possibly incomplete) contractions}\right],
\end{eqnarray}
where a contraction is defined as
\begin{equation}
  C[\phi_I(x)\phi_I^+(y)] \equiv B_F(x-y). 
\end{equation} 
Note that for Brownian processes, only \textsl{future directed}
contractions contribute in the above expression, because we find from (\ref{retprop}) and (\ref{feynprop})
that $B_F=B_R$.
As an example, assume the following string of fields is already
time-ordered, then
\begin{eqnarray}\label{correlexample}
  T\left\{\phi_I(a) \phi_I^+(b) \phi_I(c) \phi_I^+(d)\right\}
  &=& N\left[\phi_I^+(b) \phi_I^+(d) \phi_I(a) \phi_I(c)\right]\nonumber\\
  & & + B_F(a,b)N\left[\phi_I(c)\phi_I^+(d)\right]\nonumber\\
  & & + B_F(a,d)N\left[\phi_I(b)^+\phi_I(c)\right]\nonumber\\
  & & + B_F(c,d)N\left[\phi_I(a)\phi_I^+(b)\right]\nonumber\\
  & & + B_F(a,b) B_F(c,d).
\end{eqnarray}
\noindent A proof of Wick's theorem can be found in \cite{Peskin}.
Representing the propagator $B_F(x,y)$ by a directed line joining the
points $x$ and $y$,
\begin{center}
\setlength{\unitlength}{1973sp}%
\begingroup\makeatletter\ifx\SetFigFont\undefined%
\gdef\SetFigFont#1#2#3#4#5{%
  \reset@font\fontsize{#1}{#2pt}%
  \fontfamily{#3}\fontseries{#4}\fontshape{#5}%
  \selectfont}%
\fi\endgroup%
\begin{picture}(2850,489)(826,-733)
\thinlines
\put(901,-361){\vector( 1, 0){1200}}
\put(2101,-361){\line( 1, 0){900}}
\put(3676,-436){\makebox(0,0)[lb]{\smash{\SetFigFont{8}{9.6}{\rmdefault}{\mddefault}{\updefault}
$ = B_F(x,y)$
}}}
\put(2926,-661){\makebox(0,0)[lb]{\smash{\SetFigFont{8}{9.6}{\rmdefault}{\mddefault}{\updefault}
$y$
}}}
\put(826,-661){\makebox(0,0)[lb]{\smash{\SetFigFont{8}{9.6}{\rmdefault}{\mddefault}{\updefault}
$x$
}}}
\end{picture}
\end{center}
we can write correlation functions
in diagrammatic form. Note that the normal ordered terms will 
not contribute to the correlation function, as the sandwiching between
vacuum states annihilates them. The integral over $\mathcal{H}_I$ in (\ref{corrfunc})
will lead to diagrams with vertices, where all occurring vertices are
integrated over. 

Before we undertake the enterprise of calculating correlation
functions for a particular theory, however, we can use Wick's theorem
to simplify expression (\ref{corrfunc}) significantly. As only
future-directed contractions survive, we immediately have 
\begin{equation}
  \bra{\Omega}  T \exp(-I\int d^{d+1}z \mathcal{H}_I(z)) \ket{\Omega} = 1,  
\end{equation}
and hence formula (\ref{corrfunc}) for the computation of correlation functions of the
interacting theory simplifies to 
\begin{equation}\label{Feynseries}
  \bra{\Omega} T\left\{ \phi_I(x_1) \dots \phi_I(x_n) \exp(-I\int
          d^{d+1}z \mathcal{H}_I(z)) \right\} \ket{\Omega}.
\end{equation}
This is consistent with a general result for any interacting quantum field
theory, namely  that the denominator in
(\ref{corrfunc}) cancels all disconnected vacuum diagrams in the
numerator (see e.g. \cite{Peskin}). In the present case of Brownian
motion, however, we saw above that the 
Feynman propagator is retarded, and so
there simply are no non-vanishing vacuum bubbles. 

For general interactions $\mathcal{H}_I$, it is non-trivial to
calculate the $(n,m)$-point correlation functions, and very often it
is impossible to do so exactly. In those cases, which present the rule rather than the
exception in realistic quantum field theories, one must resort to a
truncated series expansion of (\ref{Feynseries}).
The diagrammatic representation of a summand in such a perturbation series
is called a Feynman diagram.

In the next section, we will show that the binary branching and dying model is
exactly solvable. Nevertheless, we will first expand the corresponding
correlation functions, and then exactly sum the resulting series by
integration of a set of recurrence relations.

%

\section{Example: Binary Branching Brownian Motion}\label{sec_bnd}
\subsection{Feynman Rules}
Binary branching and dying processes of Brownian particles are
easily formulated as interaction terms in the field theoretical
treatment of Brownian motion. This model is completely specified by
the Lagrangian density  
\begin{equation}\label{bnd}
  \mathcal{L} = \mathcal{L}_B[\phi] + \gamma \alpha \phi^* + \gamma \beta \phi^* \phi \phi,
\end{equation} 
where $\gamma \in \reals^+_0$, $\alpha, \beta \in [0,1]$, and $\alpha
+ \beta = 1$. 
The fertility distribution $\{p_k: k=0,1,2\}$, encoding the
probability that a dying particle will give birth to $k$ children,
will turn out to be $p=(\alpha, 0, \beta)$. The parameter $\gamma$
will be identified below as the clock rate of the exponential clock
controlling the process. Note, that the equation of motion corresponding to
$\mathcal{L}$ is the reaction diffusion equation (\ref{branching}) with 
$p=(\alpha,0,\beta)$ for classical binary branching Brownian motion.
The free theory $\mathcal{L}_B$ is as defined in (\ref{clockmodel}).
Applying the analysis of the previous section to the model
(\ref{bnd}), one obtains the position space Feynman rules \cite{Peskin}\\
\begin{center}
\setlength{\unitlength}{1973sp}%
\begingroup\makeatletter\ifx\SetFigFont\undefined%
\gdef\SetFigFont#1#2#3#4#5{%
  \reset@font\fontsize{#1}{#2pt}%
  \fontfamily{#3}\fontseries{#4}\fontshape{#5}%
  \selectfont}%
\fi\endgroup%
\begin{picture}(2790,3544)(661,-3430)
\thinlines
\put(1883,-3061){\circle{140}}
\put(901,-61){\vector( 1, 0){600}}
\put(1501,-61){\line( 1, 0){600}}
\put(901,-1561){\vector( 1, 0){300}}
\put(1201,-1561){\line( 1, 0){300}}
\multiput(1501,-1561)(12.00000,12.00000){26}{\makebox(3.3333,23.3333){\SetFigFont{5}{6}{\rmdefault}{\mddefault}{\updefault}.}}
\put(1801,-1261){\vector( 1, 1){0}}
\multiput(1801,-1261)(12.00000,12.00000){26}{\makebox(3.3333,23.3333){\SetFigFont{5}{6}{\rmdefault}{\mddefault}{\updefault}.}}
\multiput(1501,-1561)(12.00000,-12.00000){26}{\makebox(3.3333,23.3333){\SetFigFont{5}{6}{\rmdefault}{\mddefault}{\updefault}.}}
\put(1801,-1861){\vector( 1,-1){0}}
\multiput(1801,-1861)(12.00000,-12.00000){26}{\makebox(3.3333,23.3333){\SetFigFont{5}{6}{\rmdefault}{\mddefault}{\updefault}.}}
\put(901,-3061){\vector( 1, 0){600}}
\put(1201,-3061){\line( 1, 0){600}}
\multiput(1836,-3023)(12.14286,-12.14286){8}{\makebox(3.3333,23.3333){\SetFigFont{5}{6}{\rmdefault}{\mddefault}{\updefault}.}}
\multiput(1931,-3018)(-12.81250,-12.81250){9}{\makebox(3.3333,23.3333){\SetFigFont{5}{6}{\rmdefault}{\mddefault}{\updefault}.}}
\put(3451,-1711){\makebox(0,0)[lb]{\smash{\SetFigFont{12}{14.4}{\rmdefault}{\mddefault}{\itdefault}
}}}
\put(2829,-174){\makebox(0,0)[lb]{\smash{\SetFigFont{12}{14.4}{\familydefault}{\mddefault}{\updefault}
$=    B_F(x,y)$
}}}
\put(2844,-1665){\makebox(0,0)[lb]{\smash{\SetFigFont{12}{14.4}{\familydefault}{\mddefault}{\updefault}
$= I \gamma \beta \int d^{d+1} z$
}}}
\put(2828,-3130){\makebox(0,0)[lb]{\smash{\SetFigFont{12}{14.4}{\familydefault}{\mddefault}{\updefault}
$ = I \gamma \alpha \int d^{d+1} z$
}}}
\put(1636,-3430){\makebox(0,0)[lb]{\smash{\SetFigFont{12}{14.4}{\familydefault}{\mddefault}{\updefault}
$z$
}}}
\put(1223,-1915){\makebox(0,0)[lb]{\smash{\SetFigFont{12}{14.4}{\familydefault}{\mddefault}{\updefault}
$z$
}}}
\put(661,-390){\makebox(0,0)[lb]{\smash{\SetFigFont{12}{14.4}{\familydefault}{\mddefault}{\updefault}
$x$
}}}
\put(1913,-406){\makebox(0,0)[lb]{\smash{\SetFigFont{12}{14.4}{\familydefault}{\mddefault}{\updefault}
$y$
}}}
\end{picture}
\end{center}
as the fundamental building blocks of the trees generated by the
branching process. Thus, at last, we recover the trees of branching
Brownian motion from the field theoretical formulation! It is an
interesting question of whether and how one can extract information on
the genealogy of the process from the Feynman series, or indeed the
exact summability shown below. The development of such ideas, however,
is beyond the scope of the present paper. 
Before studying more complicated questions within the binary branching model, we calculate the probability that a single particle
will either branch or die within a specified time interval.

\subsection{Exponential Clock}
The amplitude for a particle starting at spacetime point $x$ to die within a time interval $\Delta\tau$
is given by
\begin{equation}
  I \gamma\alpha \int_{x^0}^{x^0 +\Delta\tau} dy^0 \int d^dy B_F(x,y) = I\alpha\left(1-e^{-\gamma\Delta\tau}\right), 
\end{equation}
using the Feynman rules above. Similarly, one obtains the probability
for a branching event within the time interval $\Delta\tau$. Hence
the total probability for the occurrence of branching or dying within
$\Delta\tau$ is found by the $\Gamma_+$ projection as 
\begin{equation}\label{expclock}
  1-e^{-\gamma\Delta\tau}.
\end{equation}
This exponential clock behaviour is crucial for the
process being of Markov type, as explained in section {\ref{sec_review}}. In
the field theoretical treatment, it need not be imposed, but is
\textsl{enforced}, unless $\gamma=0$ in (\ref{bnd}), but then all
interaction is switched off and one is left with free Brownian
motion. Hence, the description of binary branching by the Lagrangian
density (\ref{bnd}), within the field theoretical framework, implicitly
contains the branching tree, the exponential clock, and the fertility
distribution, which must all be separately specified in the
conventional approach. 

\subsection{Exact summation of the binary branching Brownian motion}
The fully interacting model (\ref{bnd}) provides a non-trivial test bed
for our formalism. Here, the one-point function already
has contributions from any order in $\alpha$ and $\beta$,
\begin{center}
\setlength{\unitlength}{1973sp}%
\begingroup\makeatletter\ifx\SetFigFont\undefined%
\gdef\SetFigFont#1#2#3#4#5{%
  \reset@font\fontsize{#1}{#2pt}%
  \fontfamily{#3}\fontseries{#4}\fontshape{#5}%
  \selectfont}%
\fi\endgroup%
\begin{picture}(10596,857)(109,-625)
\thinlines
\put(5963,-23){\circle{108}}
\put(4162,-24){\circle{108}}
\put(6337,-24){\circle{108}}
\put(8150,171){\circle{108}}
\put(8157,-219){\circle{108}}
\put(9747,150){\circle{108}}
\put(9732,-233){\circle{108}}
\put(10414,165){\circle{108}}
\put(10429,-225){\circle{108}}
\put(4163,-23){\circle{108}}
\put(3053,-363){\circle{108}}
\put(4853,-363){\circle{108}}
\put(6647,-363){\circle{108}}
\put(8440,-358){\circle{108}}
\put(10251,-363){\circle{108}}
\put(1101,-356){\circle{524}}
\put(1801,-361){\vector( 1, 0){600}}
\put(2401,-361){\line( 1, 0){600}}
\put(3601,-361){\vector( 1, 0){600}}
\put(4201,-361){\line( 1, 0){600}}
\multiput(3901,-361)(10.22727,13.63636){23}{\makebox(3.3333,23.3333){\SetFigFont{5}{6}{\rmdefault}{\mddefault}{\updefault}.}}
\put(5401,-361){\vector( 1, 0){600}}
\put(6001,-361){\line( 1, 0){600}}
\multiput(5701,-361)(10.22727,13.63636){23}{\makebox(3.3333,23.3333){\SetFigFont{5}{6}{\rmdefault}{\mddefault}{\updefault}.}}
\multiput(6076,-361)(10.22727,13.63636){23}{\makebox(3.3333,23.3333){\SetFigFont{5}{6}{\rmdefault}{\mddefault}{\updefault}.}}
\put(7201,-361){\vector( 1, 0){600}}
\put(7801,-361){\line( 1, 0){600}}
\multiput(7501,-361)(10.22727,13.63636){23}{\makebox(3.3333,23.3333){\SetFigFont{5}{6}{\rmdefault}{\mddefault}{\updefault}.}}
\put(7726,-61){\line( 5, 3){375}}
\multiput(7726,-61)(15.62500,-6.25000){25}{\makebox(3.3333,23.3333){\SetFigFont{5}{6}{\rmdefault}{\mddefault}{\updefault}.}}
\put(9001,-361){\vector( 1, 0){600}}
\put(9601,-361){\line( 1, 0){600}}
\multiput(9313,-70)(-10.22727,-13.63636){23}{\makebox(3.3333,23.3333){\SetFigFont{5}{6}{\rmdefault}{\mddefault}{\updefault}.}}
\multiput(9309,-79)(15.62500,-6.25000){25}{\makebox(3.3333,23.3333){\SetFigFont{5}{6}{\rmdefault}{\mddefault}{\updefault}.}}
\put(9319,-68){\line( 5, 3){375}}
\multiput(9994,-43)(-10.22727,-13.63636){23}{\makebox(3.3333,23.3333){\SetFigFont{5}{6}{\rmdefault}{\mddefault}{\updefault}.}}
\multiput(9991,-60)(15.62500,-6.25000){25}{\makebox(3.3333,23.3333){\SetFigFont{5}{6}{\rmdefault}{\mddefault}{\updefault}.}}
\put(9987,-62){\line( 5, 3){375}}
\multiput(4126,  4)(12.50000,-12.50000){7}{\makebox(3.3333,23.3333){\SetFigFont{5}{6}{\rmdefault}{\mddefault}{\updefault}.}}
\multiput(5928, 12)(12.50000,-12.50000){7}{\makebox(3.3333,23.3333){\SetFigFont{5}{6}{\rmdefault}{\mddefault}{\updefault}.}}
\multiput(6303, 12)(12.50000,-12.50000){7}{\makebox(3.3333,23.3333){\SetFigFont{5}{6}{\rmdefault}{\mddefault}{\updefault}.}}
\multiput(8112,206)(12.50000,-12.50000){7}{\makebox(3.3333,23.3333){\SetFigFont{5}{6}{\rmdefault}{\mddefault}{\updefault}.}}
\multiput(8122,-179)(12.50000,-12.50000){7}{\makebox(3.3333,23.3333){\SetFigFont{5}{6}{\rmdefault}{\mddefault}{\updefault}.}}
\multiput(9717,187)(12.50000,-12.50000){7}{\makebox(3.3333,23.3333){\SetFigFont{5}{6}{\rmdefault}{\mddefault}{\updefault}.}}
\multiput(9710,-206)(12.50000,-12.50000){7}{\makebox(3.3333,23.3333){\SetFigFont{5}{6}{\rmdefault}{\mddefault}{\updefault}.}}
\multiput(10385,204)(12.50000,-12.50000){7}{\makebox(3.3333,23.3333){\SetFigFont{5}{6}{\rmdefault}{\mddefault}{\updefault}.}}
\multiput(10395,-191)(12.50000,-12.50000){7}{\makebox(3.3333,23.3333){\SetFigFont{5}{6}{\rmdefault}{\mddefault}{\updefault}.}}
\multiput(4200,  9)(-12.50000,-12.50000){7}{\makebox(3.3333,23.3333){\SetFigFont{5}{6}{\rmdefault}{\mddefault}{\updefault}.}}
\multiput(6010,  9)(-12.50000,-12.50000){7}{\makebox(3.3333,23.3333){\SetFigFont{5}{6}{\rmdefault}{\mddefault}{\updefault}.}}
\multiput(6385,  9)(-12.50000,-12.50000){7}{\makebox(3.3333,23.3333){\SetFigFont{5}{6}{\rmdefault}{\mddefault}{\updefault}.}}
\multiput(8181,198)(-12.50000,-12.50000){7}{\makebox(3.3333,23.3333){\SetFigFont{5}{6}{\rmdefault}{\mddefault}{\updefault}.}}
\multiput(8191,-187)(-12.50000,-12.50000){7}{\makebox(3.3333,23.3333){\SetFigFont{5}{6}{\rmdefault}{\mddefault}{\updefault}.}}
\multiput(9786,183)(-12.50000,-12.50000){7}{\makebox(3.3333,23.3333){\SetFigFont{5}{6}{\rmdefault}{\mddefault}{\updefault}.}}
\multiput(9776,-202)(-12.50000,-12.50000){7}{\makebox(3.3333,23.3333){\SetFigFont{5}{6}{\rmdefault}{\mddefault}{\updefault}.}}
\multiput(10451,198)(-12.50000,-12.50000){7}{\makebox(3.3333,23.3333){\SetFigFont{5}{6}{\rmdefault}{\mddefault}{\updefault}.}}
\multiput(10466,-192)(-12.50000,-12.50000){7}{\makebox(3.3333,23.3333){\SetFigFont{5}{6}{\rmdefault}{\mddefault}{\updefault}.}}
\multiput(3090,-406)(-11.41463,14.26829){7}{\makebox(3.3333,23.3333){\SetFigFont{5}{6}{\rmdefault}{\mddefault}{\updefault}.}}
\multiput(3082,-331)(-12.62296,-15.14755){6}{\makebox(3.3333,23.3333){\SetFigFont{5}{6}{\rmdefault}{\mddefault}{\updefault}.}}
\multiput(4890,-398)(-12.16667,12.16667){7}{\makebox(3.3333,23.3333){\SetFigFont{5}{6}{\rmdefault}{\mddefault}{\updefault}.}}
\multiput(4878,-323)(-11.90400,-15.87200){6}{\makebox(3.3333,23.3333){\SetFigFont{5}{6}{\rmdefault}{\mddefault}{\updefault}.}}
\multiput(6620,-323)(11.90400,-15.87200){6}{\makebox(3.3333,23.3333){\SetFigFont{5}{6}{\rmdefault}{\mddefault}{\updefault}.}}
\multiput(6688,-331)(-10.79235,-12.95082){7}{\makebox(3.3333,23.3333){\SetFigFont{5}{6}{\rmdefault}{\mddefault}{\updefault}.}}
\multiput(8408,-331)(12.80000,-12.80000){6}{\makebox(3.3333,23.3333){\SetFigFont{5}{6}{\rmdefault}{\mddefault}{\updefault}.}}
\multiput(8472,-327)(-13.90000,-13.90000){6}{\makebox(3.3333,23.3333){\SetFigFont{5}{6}{\rmdefault}{\mddefault}{\updefault}.}}
\multiput(10223,-327)(11.10568,-13.88210){7}{\makebox(3.3333,23.3333){\SetFigFont{5}{6}{\rmdefault}{\mddefault}{\updefault}.}}
\multiput(10287,-331)(-12.16667,-12.16667){7}{\makebox(3.3333,23.3333){\SetFigFont{5}{6}{\rmdefault}{\mddefault}{\updefault}.}}
\put(631,-361){\line( 1, 0){205}}
\put(121,-361){\vector( 1, 0){505}}
\put(3226,-436){\makebox(0,0)[lb]{\smash{\SetFigFont{8}{9.6}{\familydefault}{\mddefault}{\updefault}
+
}}}
\put(3226,-436){\makebox(0,0)[lb]{\smash{\SetFigFont{8}{9.6}{\familydefault}{\mddefault}{\updefault}
+
}}}
\put(5026,-436){\makebox(0,0)[lb]{\smash{\SetFigFont{8}{9.6}{\familydefault}{\mddefault}{\updefault}
+
}}}
\put(6826,-436){\makebox(0,0)[lb]{\smash{\SetFigFont{8}{9.6}{\familydefault}{\mddefault}{\updefault}
+
}}}
\put(8626,-436){\makebox(0,0)[lb]{\smash{\SetFigFont{8}{9.6}{\familydefault}{\mddefault}{\updefault}
+
}}}
\put(10441,-428){\makebox(0,0)[lb]{\smash{\SetFigFont{8}{9.6}{\familydefault}{\mddefault}{\updefault}
+
}}}
\put(10705,-376){\makebox(0,0)[lb]{\smash{\SetFigFont{8}{9.6}{\familydefault}{\mddefault}{\updefault}
...
}}}
\put(1546,-411){\makebox(0,0)[lb]{\smash{\SetFigFont{6}{7.2}{\rmdefault}{\mddefault}{\updefault}
=
}}}
\put(966,-421){\makebox(0,0)[lb]{\smash{\SetFigFont{6}{7.2}{\familydefault}{\mddefault}{\updefault}
$A_t$
}}}
\end{picture},
\end{center}
where the subscript $t$ denotes the earliest time at which all
particles occurring in the series have died.
Similarly, the two-point correlation function is
\begin{center}
\setlength{\unitlength}{1973sp}%
\begingroup\makeatletter\ifx\SetFigFont\undefined%
\gdef\SetFigFont#1#2#3#4#5{%
  \reset@font\fontsize{#1}{#2pt}%
  \fontfamily{#3}\fontseries{#4}\fontshape{#5}%
  \selectfont}%
\fi\endgroup%
\begin{picture}(10741,1011)(109,-628)
\thinlines
\put(5963,-23){\circle{108}}
\put(4162,-24){\circle{108}}
\put(6337,-24){\circle{108}}
\put(8150,171){\circle{108}}
\put(8157,-219){\circle{108}}
\put(9732,-233){\circle{108}}
\put(10414,165){\circle{108}}
\put(10429,-225){\circle{108}}
\put(4163,-23){\circle{108}}
\put(822,-359){\circle{524}}
\put(10664,322){\circle{108}}
\put(1801,-361){\vector( 1, 0){600}}
\put(2401,-361){\line( 1, 0){600}}
\put(3601,-361){\vector( 1, 0){600}}
\put(4201,-361){\line( 1, 0){600}}
\multiput(3901,-361)(10.22727,13.63636){23}{\makebox(3.3333,23.3333){\SetFigFont{5}{6}{\rmdefault}{\mddefault}{\updefault}.}}
\put(5401,-361){\vector( 1, 0){600}}
\put(6001,-361){\line( 1, 0){600}}
\multiput(5701,-361)(10.22727,13.63636){23}{\makebox(3.3333,23.3333){\SetFigFont{5}{6}{\rmdefault}{\mddefault}{\updefault}.}}
\multiput(6076,-361)(10.22727,13.63636){23}{\makebox(3.3333,23.3333){\SetFigFont{5}{6}{\rmdefault}{\mddefault}{\updefault}.}}
\put(7201,-361){\vector( 1, 0){600}}
\multiput(7501,-361)(10.22727,13.63636){23}{\makebox(3.3333,23.3333){\SetFigFont{5}{6}{\rmdefault}{\mddefault}{\updefault}.}}
\put(7726,-61){\line( 5, 3){375}}
\multiput(7726,-61)(15.62500,-6.25000){25}{\makebox(3.3333,23.3333){\SetFigFont{5}{6}{\rmdefault}{\mddefault}{\updefault}.}}
\put(9001,-361){\vector( 1, 0){600}}
\put(9601,-361){\line( 1, 0){1219}}
\multiput(9313,-70)(-10.22727,-13.63636){23}{\makebox(3.3333,23.3333){\SetFigFont{5}{6}{\rmdefault}{\mddefault}{\updefault}.}}
\multiput(9309,-79)(15.62500,-6.25000){25}{\makebox(3.3333,23.3333){\SetFigFont{5}{6}{\rmdefault}{\mddefault}{\updefault}.}}
\multiput(9994,-43)(-10.22727,-13.63636){23}{\makebox(3.3333,23.3333){\SetFigFont{5}{6}{\rmdefault}{\mddefault}{\updefault}.}}
\multiput(9991,-60)(15.62500,-6.25000){25}{\makebox(3.3333,23.3333){\SetFigFont{5}{6}{\rmdefault}{\mddefault}{\updefault}.}}
\put(9987,-62){\line( 5, 3){375}}
\multiput(4126,  4)(12.50000,-12.50000){7}{\makebox(3.3333,23.3333){\SetFigFont{5}{6}{\rmdefault}{\mddefault}{\updefault}.}}
\multiput(5928, 12)(12.50000,-12.50000){7}{\makebox(3.3333,23.3333){\SetFigFont{5}{6}{\rmdefault}{\mddefault}{\updefault}.}}
\multiput(6303, 12)(12.50000,-12.50000){7}{\makebox(3.3333,23.3333){\SetFigFont{5}{6}{\rmdefault}{\mddefault}{\updefault}.}}
\multiput(8112,206)(12.50000,-12.50000){7}{\makebox(3.3333,23.3333){\SetFigFont{5}{6}{\rmdefault}{\mddefault}{\updefault}.}}
\multiput(8122,-179)(12.50000,-12.50000){7}{\makebox(3.3333,23.3333){\SetFigFont{5}{6}{\rmdefault}{\mddefault}{\updefault}.}}
\multiput(9710,-206)(12.50000,-12.50000){7}{\makebox(3.3333,23.3333){\SetFigFont{5}{6}{\rmdefault}{\mddefault}{\updefault}.}}
\multiput(10385,204)(12.50000,-12.50000){7}{\makebox(3.3333,23.3333){\SetFigFont{5}{6}{\rmdefault}{\mddefault}{\updefault}.}}
\multiput(10395,-191)(12.50000,-12.50000){7}{\makebox(3.3333,23.3333){\SetFigFont{5}{6}{\rmdefault}{\mddefault}{\updefault}.}}
\multiput(4200,  9)(-12.50000,-12.50000){7}{\makebox(3.3333,23.3333){\SetFigFont{5}{6}{\rmdefault}{\mddefault}{\updefault}.}}
\multiput(6010,  9)(-12.50000,-12.50000){7}{\makebox(3.3333,23.3333){\SetFigFont{5}{6}{\rmdefault}{\mddefault}{\updefault}.}}
\multiput(6385,  9)(-12.50000,-12.50000){7}{\makebox(3.3333,23.3333){\SetFigFont{5}{6}{\rmdefault}{\mddefault}{\updefault}.}}
\multiput(8181,198)(-12.50000,-12.50000){7}{\makebox(3.3333,23.3333){\SetFigFont{5}{6}{\rmdefault}{\mddefault}{\updefault}.}}
\multiput(8191,-187)(-12.50000,-12.50000){7}{\makebox(3.3333,23.3333){\SetFigFont{5}{6}{\rmdefault}{\mddefault}{\updefault}.}}
\multiput(9776,-202)(-12.50000,-12.50000){7}{\makebox(3.3333,23.3333){\SetFigFont{5}{6}{\rmdefault}{\mddefault}{\updefault}.}}
\multiput(10451,198)(-12.50000,-12.50000){7}{\makebox(3.3333,23.3333){\SetFigFont{5}{6}{\rmdefault}{\mddefault}{\updefault}.}}
\multiput(10466,-192)(-12.50000,-12.50000){7}{\makebox(3.3333,23.3333){\SetFigFont{5}{6}{\rmdefault}{\mddefault}{\updefault}.}}
\put(121,-361){\vector( 1, 0){245}}
\put(351,-361){\line( 1, 0){205}}
\put(1086,-361){\vector( 1, 0){245}}
\put(1271,-361){\line( 1, 0){205}}
\put(7801,-361){\line( 1, 0){600}}
\put(9319,-68){\line( 3, 1){1282.800}}
\multiput(10632,354)(12.50000,-12.50000){7}{\makebox(3.3333,23.3333){\SetFigFont{5}{6}{\rmdefault}{\mddefault}{\updefault}.}}
\multiput(10707,349)(-12.50000,-12.50000){7}{\makebox(3.3333,23.3333){\SetFigFont{5}{6}{\rmdefault}{\mddefault}{\updefault}.}}
\put(3226,-436){\makebox(0,0)[lb]{\smash{\SetFigFont{8}{9.6}{\familydefault}{\mddefault}{\updefault}
+
}}}
\put(3226,-436){\makebox(0,0)[lb]{\smash{\SetFigFont{8}{9.6}{\familydefault}{\mddefault}{\updefault}
+
}}}
\put(5026,-436){\makebox(0,0)[lb]{\smash{\SetFigFont{8}{9.6}{\familydefault}{\mddefault}{\updefault}
+
}}}
\put(6826,-436){\makebox(0,0)[lb]{\smash{\SetFigFont{8}{9.6}{\familydefault}{\mddefault}{\updefault}
+
}}}
\put(8626,-436){\makebox(0,0)[lb]{\smash{\SetFigFont{8}{9.6}{\familydefault}{\mddefault}{\updefault}
+
}}}
\put(1546,-411){\makebox(0,0)[lb]{\smash{\SetFigFont{6}{7.2}{\rmdefault}{\mddefault}{\updefault}
=
}}}
\put(691,-411){\makebox(0,0)[lb]{\smash{\SetFigFont{6}{7.2}{\familydefault}{\mddefault}{\updefault}
$D$
}}}
\put(10850,-381){\makebox(0,0)[lb]{\smash{\SetFigFont{8}{9.6}{\familydefault}{\mddefault}{\updefault}
...
}}}
\end{picture}
\end{center}
with contributions of order $\mathcal{O}(\alpha) = \mathcal{O}(\beta)$.\\
We will put to use the special properties of the model at hand to exactly sum these Feyman series. The one-point function $A(z)$ and the two-point function
$D(x,y)$ of the interacting theory satisfy the coupled
recurrence relations
\begin{center}
\setlength{\unitlength}{2763sp}%
\begingroup\makeatletter\ifx\SetFigFont\undefined%
\gdef\SetFigFont#1#2#3#4#5{%
  \reset@font\fontsize{#1}{#2pt}%
  \fontfamily{#3}\fontseries{#4}\fontshape{#5}%
  \selectfont}%
\fi\endgroup%
\begin{picture}(9075,3877)(751,-3158)
\thinlines
\put(5776,-361){\circle{150}}
\put(2791,-391){\circle{778}}
\put(8557,322){\circle{778}}
\put(8564,-1036){\circle{778}}
\put(7492,-1928){\circle{778}}
\put(8407,-2761){\circle{778}}
\put(1956,-2753){\circle{778}}
\put(4201,-361){\line( 1, 0){1500}}
\put(901,-361){\line( 1, 0){1500}}
\put(6901,-361){\line( 1, 0){900}}
\put(7801,-361){\line( 1, 1){450}}
\put(7801,-361){\line( 1,-1){450}}
\put(901,-2761){\line( 1, 0){665}}
\put(2346,-2761){\line( 1, 0){655}}
\put(4201,-2761){\line( 1, 0){1500}}
\put(6901,-2761){\line( 1, 0){1116}}
\put(8797,-2761){\line( 1, 0){504}}
\put(7501,-2761){\line( 0, 1){450}}
\put(3436,-421){\makebox(0,0)[lb]{\smash{\SetFigFont{10}{12.0}{\familydefault}{\mddefault}{\updefault}
=
}}}
\put(6301,-436){\makebox(0,0)[lb]{\smash{\SetFigFont{10}{12.0}{\familydefault}{\mddefault}{\updefault}
+
}}}
\put(7621,-556){\makebox(0,0)[lb]{\smash{\SetFigFont{8}{9.6}{\familydefault}{\mddefault}{\updefault}
$w$
}}}
\put(766,-601){\makebox(0,0)[lb]{\smash{\SetFigFont{8}{9.6}{\familydefault}{\mddefault}{\updefault}
$z$
}}}
\put(4096,-646){\makebox(0,0)[lb]{\smash{\SetFigFont{8}{9.6}{\familydefault}{\mddefault}{\updefault}
$z$
}}}
\put(4096,-646){\makebox(0,0)[lb]{\smash{\SetFigFont{8}{9.6}{\familydefault}{\mddefault}{\updefault}
$z$
}}}
\put(6781,-646){\makebox(0,0)[lb]{\smash{\SetFigFont{8}{9.6}{\familydefault}{\mddefault}{\updefault}
$z$
}}}
\put(3376,-2836){\makebox(0,0)[lb]{\smash{\SetFigFont{10}{12.0}{\familydefault}{\mddefault}{\updefault}
=
}}}
\put(6301,-2836){\makebox(0,0)[lb]{\smash{\SetFigFont{10}{12.0}{\familydefault}{\mddefault}{\updefault}
+
}}}
\put(751,-3061){\makebox(0,0)[lb]{\smash{\SetFigFont{8}{9.6}{\familydefault}{\mddefault}{\updefault}
$x$
}}}
\put(2926,-3061){\makebox(0,0)[lb]{\smash{\SetFigFont{8}{9.6}{\familydefault}{\mddefault}{\updefault}
$y$
}}}
\put(5626,-3061){\makebox(0,0)[lb]{\smash{\SetFigFont{8}{9.6}{\familydefault}{\mddefault}{\updefault}
$y$
}}}
\put(9226,-3061){\makebox(0,0)[lb]{\smash{\SetFigFont{8}{9.6}{\familydefault}{\mddefault}{\updefault}
$y$
}}}
\put(4051,-3061){\makebox(0,0)[lb]{\smash{\SetFigFont{8}{9.6}{\familydefault}{\mddefault}{\updefault}
$x$
}}}
\put(6826,-3061){\makebox(0,0)[lb]{\smash{\SetFigFont{8}{9.6}{\familydefault}{\mddefault}{\updefault}
$x$
}}}
\put(7351,-3061){\makebox(0,0)[lb]{\smash{\SetFigFont{8}{9.6}{\familydefault}{\mddefault}{\updefault}
$w$
}}}
\put(9826,-436){\makebox(0,0)[lb]{\smash{\SetFigFont{8}{9.6}{\familydefault}{\mddefault}{\updefault}
$\label{rec1}$
}}}
\put(9826,-2836){\makebox(0,0)[lb]{\smash{\SetFigFont{8}{9.6}{\familydefault}{\mddefault}{\updefault}
$\label{rec2}$
}}}
\put(1846,-2821){\makebox(0,0)[lb]{\smash{\SetFigFont{7}{8.4}{\familydefault}{\mddefault}{\updefault}
$D$
}}}
\put(8281,-2791){\makebox(0,0)[lb]{\smash{\SetFigFont{7}{8.4}{\familydefault}{\mddefault}{\updefault}
$D$
}}}
\put(2536,-428){\makebox(0,0)[lb]{\smash{\SetFigFont{7}{8.4}{\familydefault}{\mddefault}{\updefault}
$A_t(z)$
}}}
\put(8288,299){\makebox(0,0)[lb]{\smash{\SetFigFont{7}{8.4}{\familydefault}{\mddefault}{\updefault}
$A_t(w)$
}}}
\put(8303,-1089){\makebox(0,0)[lb]{\smash{\SetFigFont{7}{8.4}{\familydefault}{\mddefault}{\updefault}
$A_t(w)$
}}}
\put(7150,-1981){\makebox(0,0)[lb]{\smash{\SetFigFont{7}{8.4}{\familydefault}{\mddefault}{\updefault}
$A_{y^0}(w)$
}}}
\end{picture}
\end{center}
\noindent Algebraically, these read 
\begin{eqnarray}
A_t(z) \hspace{-5pt} &=& \hspace{-5pt} I\gamma\alpha \int_{z_0}^t dw^0 \int d^d w B_F(z,w) +
I \gamma\beta
\int_{z^0}^t dw^0 \int d^d w \,B_F(z,w) A_t^2(w),\label{Aintegral}\nonumber\\
D(x,y) \hspace{-5pt} &=& \hspace{-5pt}B_F(x,y) + I\,\gamma\beta\, \int_{x^0}^{y^0} dw^0 \int d^dw \,
B_F(x,w) A_{y_0}(w) D(w,y).\label{Dintegral}\nonumber
\end{eqnarray}

The translational invariance of the theory implies that the one-point
function $A_t(z)$ only depends on the difference $\tau \equiv t -
z^0$, and hence we denote $A_t(z)$ as $A(t-z^0)$. The recurrence relation
for $A_t$ can then be written
\begin{equation}\label{Aintegral2}
  A(\tau) = I\gamma\alpha \int_0^\tau dw^0 e^{-\gamma w^0} + I
  \gamma\beta \int_0^\tau dw^0 e^{-\gamma w^0} A^2(\tau-w^0), 
\end{equation} 
using that the spatial integral over $B_F$ is normalized to
unity. Multiplication of (\ref{Aintegral2}) by $e^{\gamma \tau}$
renders the second integrand independent of $\tau$. By differentiation
with respect to $\tau$ we obtain the differential form of the recurrence
relation (\ref{Aintegral}). Writing $\tilde A$ for the projection
$\Gamma_+ A$, it reads
\begin{equation}
  \frac{d\tilde A(\tau)}{d\tau} = \gamma\alpha -\gamma \tilde A(\tau) + \gamma\beta
  \tilde A^2(\tau),\qquad \tilde A(0)=0.\label{ode}
\end{equation}
Indeed, equation (\ref{ode}) is equal to (\ref{dualN}) for $\theta=0$,
because there we have $W(\varphi) = \alpha + \beta \varphi^2$ for a binary
branching mechanism.
The unique solution of (\ref{ode}) subject to the constraint
$\alpha+\beta = 1$ is 
\begin{equation}
  \tilde A(\tau) = \left\{\begin{array}{l}
    1-\frac{2}{\gamma\tau+2}, \qquad \textrm{iff} \quad \alpha=\beta=\frac{1}{2}\\[10pt]
    \frac{1}{2\beta} \left(
      1 - \sqrt{1-4\alpha\beta}
    \tanh\left[\frac{1}{2}\sqrt{1-4\alpha\beta} \gamma \tau + C\right]
    \right), \
    \end{array}\right. \label{Asolution} 
\end{equation}
where $\tanh(C)=(1-4\alpha\beta)^{-\frac{1}{2}}$.\\
From this result we can easily extract the \textsl{extinction probability} of
the process, i.e., the probability for the process to have died
completely at temporal infinity $\tau \rightarrow \infty$:
\begin{equation}\label{totaldying}
   \lim_{\tau\rightarrow \infty} \tilde A(\tau) =
   \frac{1-|1-2\alpha|}{2-2\alpha} =
   \left\{\begin{array}{ll}\frac{\alpha}{1-\alpha} & \textrm{for }
   \alpha < \frac{1}{2}\\ 1 & \textrm{for } \alpha \geq \frac{1}{2}\end{array}\right. 
\end{equation}
with almost sure extinction for $\alpha\geq\frac{1}{2}\geq\beta$, as one would intuitively
expect. Indeed, (\ref{totaldying}) is a classical result for branching
Brownian motion \cite{Athreya}. 
As the function $D(x,y)$ only depends
on the difference $y-x$ of the space-time points $x$ and $y$, we may
write the recurrence relation for $D(x,y)$ in the form
\begin{equation}
  D(z) = B(z) + I \gamma \beta \int_0^{z^0} dw^0 \, \int d^dw B(w) A(z^0-w^0) D(z-w). \label{dintegral2}
\end{equation}
Requiring $D$ to be integrable, standard arguments using Banach's fixpoint theorem, guarantee
existence and uniqueness of a solution to the integral equation (\ref{dintegral2}).

\subsection{Extension to Brownian catalysts}
 The extension of the model to Brownian catalysts is
straightforward in the field theoretical treatment. One simply provides
terms for both the free Brownian particle $\phi$ and the
free Brownian catalyst $\psi$, say. The type of interaction between
them is then modeled in the interaction terms. As an example, 
self-intoxicating particles $\phi$ are modeled by 
\begin{equation}
  \mathcal{L} = \mathcal{L}_B[\phi] + \mathcal{L}_B[\psi] + \gamma\alpha
  \psi^* \phi^* + \gamma\beta \phi^* \phi\psi,
\end{equation}
where $\psi$ presents the toxic substance.
The one-point function for this model is exactly solvable, now due to its simple \textsl{two-point}
function, which can be exactly summed and immediately gives the
one-point function.
The analysis of catalytic models will, in general, be more involved than
the exactly solvable binary branching or self-intoxicating model. 
However, expressing catalytic Brownian motion in the form of a
pseudo-complex field theory, most of the apparatus of quantum field
theories is applicable to these questions. This is expected to allow
insights into such processes that are hidden in the standard formalism.

\section{Conclusion}
The product structure of heat phase space emerges as the key to a geometrical
understanding of characteristic properties of Brownian
motion, most notably the non-differentiability of Brownian
paths. This structure is concisely encoded in the commutative ring of
pseudo-complex numbers, giving rise to dynamics for a pseudo-complex
valued heat amplitude.
The real-valued heat can be extracted from the heat amplitude by a unique
additive and multiplicative projection, reflecting the absence of
interference and Markov property for heat propagation. The discussion
of an abstract Hilbert module over the pseudo-complex ring provides a convenient formulation 
of one-particle Brownian motion in terms of heat amplitudes.
The appropriate kinematical framework for the discussion of Brownian
processes with non-constant particle number is the pseudo-complex Fock
space generated by the one-particle system. The operator-valued second
quantized field equations are rigorously constructed and extended by
interactions. A standard derivation of the correlation
functions for interacting theories in terms of the free fields yields
their diagrammatical representations. 

The central result of the paper is the realization that 
multi-particle Brownian processes can be conveniently studied as pseudo-complex
quantum field theories, with Brownian particles emerging as the quanta
of the heat amplitude field. Arbitrarily complex models are easily
written down in the formalism. Indeed, the
mere specification of the model dynamics, by a Lagrangian,
generates \textsl{all} elements of the conventional constructions:
The Feynman diagrams of the pseudo-complex quantum field theory
naturally generate the trees of the standard approach; interactions
cannot be switched on without also starting an exponential clock,
which guarantees the Markov property of the process; the fertility and
death rates are concisely encoded in the dynamics of the model. 
No adaptations of the formalism must be made if the dynamics are
extended by additional fields and arbitrary local interactions.
 This is in contrast to the conventional approach discussed in section
\ref{sec_review}, and indeed solves the discussed shortcomings of the latter.

The well-known binary branching Brownian motion provides
an ideal test bed for the field theoretical description, and its
one-point function is explicitly calculated, illustrating the extraction
of information in the presented formalism. The extinction probability of binary
branching Brownian motion is a point in case and yields, of course,
the classical result. 

A field theoretical treatment easily allows the
discussion of dynamical catalysts, which attract much interest in the
present probabilistic literature on the subject \cite{catalytic}. 
The application of the theory to 
quantum systems immersed in a heat bath, a question that currently
achieves a lot of attention in the context of quantum information
theory, is under investigation.


An important restriction of the systems that can be rigorously
discussed, in the context of mathematical questions, is the
requirement of exact solvability of the Dyson-Schwinger equations.
However, for phenomenological applications of branching processes to
practical problems, a diagrammatic expansion
may just be the appropriate tool for extracting information on a complex
system. 

Exerting some care in the application of standard field
theoretical results to the pseudo-complex case, much of the vast
apparatus of modern quantum field theory is now applicable to
problems on Brownian motion, and vice versa.

  


\end{document}